\documentclass[12pt,preprint]{aastex}

\newcommand{\kms}{km~s\ensuremath{^{-1}}}

\newcommand{\vlsr}{\ensuremath{v_{\rm LSR}}}
\newcommand{\hi}{\ion{H}{1}}
\newcommand{\hii}{\ion{H}{2}}
\newcommand{\hei}{\ion{He}{1}}
\newcommand{\oi}{\ion{O}{1}}

\newcommand{\sii}{\ion{S}{2}}
\newcommand{\nii}{\ion{N}{2}}
\newcommand{\ha}{H\ensuremath{\alpha}}

\newcommand{\iha}{\ensuremath{I_{\mathrm{H}\alpha}}}

\slugcomment{Astrophysical Journal Supplement, 2003, 149, in press}
\shorttitle{The WHAM Northern Sky Survey}
\shortauthors{Haffner et al.}

\begin{document}

\title{The Wisconsin H-Alpha Mapper Northern Sky Survey}

\author{L. M. Haffner, R. J. Reynolds}
\affil{Department of Astronomy, University of Wisconsin, 475 North Charter Street, Madison, WI 53706}
\email{haffner@astro.wisc.edu}
\email{reynolds@astro.wisc.edu}

\author{S. L. Tufte}
\affil{Department of Physics, Lewis and Clark College, 0615 Palatine Hill Road, MSC 15, Portland, OR 97219}
\email{tufte@lclark.edu}

\author{G. J. Madsen}
\affil{Department of Astronomy, University of Wisconsin, 475 North Charter Street, Madison, WI 53706}
\email{madsen@astro.wisc.edu}

\and

\author{K. P. Jaehnig, J. W. Percival}
\affil{Space Astronomy Lab, University of Wisconsin, 1150 University Avenue, Madison, WI 53706}
\email{kurt@sal.wisc.edu}
\email{jwp@sal.wisc.edu}

\begin{abstract}
The Wisconsin H-Alpha Mapper (WHAM) has surveyed the distribution and kinematics of  ionized gas in the Galaxy above declination $-30\arcdeg$. The WHAM Northern Sky Survey (WHAM-NSS) has an angular resolution of one degree and provides the first absolutely-calibrated, kinematically-resolved map of the \ha\ emission from the Warm Ionized Medium (WIM) within $\sim \pm100$ \kms\ of the Local Standard of Rest. Leveraging WHAM's 12 \kms\ spectral resolution, we have modeled and removed atmospheric emission and zodiacal absorption features from each of the 37,565 spectra. The resulting \ha\ profiles reveal ionized gas detected in nearly every direction on the sky with a sensitivity of 0.15 R ($3\sigma$). Complex distributions of ionized gas are revealed in the nearby spiral arms up to 1--2 kpc away from the Galactic plane. Toward the inner Galaxy, the WHAM-NSS provides information about the WIM out to the tangent point down to a few degrees from the plane. Ionized gas is also detected toward many intermediate velocity clouds at high latitudes. Several new \hii\ regions are revealed around early B-stars and evolved stellar cores (sdB/O). This work presents the details of the instrument, the survey, and the data reduction techniques. The WHAM-NSS is also presented and analyzed for its gross properties. Finally, some general conclusions are presented about the nature of the WIM as revealed by the WHAM-NSS.
\end{abstract}


\section{INTRODUCTION}

Studies of faint emission lines over the past two decades have firmly established the Warm Ionized Medium (WIM) as a significant component of the ISM, especially in the halos of disk galaxies \citep{Reynolds93,HWR99,Rand96,RD00,Dettmar98}. In our own Galaxy, measurements show the WIM to have a mass surface density about one-third that of \hi, an ionization power requirement equal to the kinetic energy from Galactic supernovae, and a characteristic scale height of about 1 kpc in the solar neighborhood \citep{Reynolds93, Ferriere01}. Although originally discovered in the 1960s through radio observations \citep{HE63,Guelin74}, information about the distribution, kinematics, and other properties of the WIM has been obtained primarily through the study of optical emission lines, particularly \ha, [\sii] $\lambda$6716, and [\nii] $\lambda$6583. Characterizing the details of the WIM is important not only for improving the picture of the ISM and the interaction among its components, but also for understanding the nature of the foreground gas through which all extragalactic and cosmological observations are made \citep[\emph{e.g.}\ ][]{Lagache+00,WMAPFG}.

Although the details of how the WIM is ionized and heated are not well understood yet, many authors have presented ideas for the source of the ionization and heating of the WIM and its relationship to other phases of the ISM. \citet{MC93}, \citet{DS94}, and \citet{DSF00} present a picture of a pervasive, warm, (originally) neutral medium being ionized by radiation from early-type stars leaking between cold \hi\ clouds or out of density bounded \hii\ regions. Potentially, such a picture results in regions of fully ionized gas and ``shadowed'' neutral gas. Others suggest that the \ha\ emission should be positively correlated with \hi\ features. \citet{Norman91} and \citet{KHR92} associate the pervasive \ha\ emission with the walls of superbubbles, ``chimneys,'' and ``worms,'' while \citet{MO77} suggest that prominent \ha\ emission is confined to the surface of neutral clouds bathed in a hot medium. \citet{SF93} and \citet{Sciama90} have suggested that the H$^+$ should be well-mixed with the neutral material, although these scenarios are inconsistent with recent measurements of the ionization fraction within the \ha\ emitting gas \citep{Reynolds+98,Hausen+02-aj}. \cite{SM00} have predicted a minimum ionizing flux present in the halo of the Milky Way contributed by old, cooling supernova remnants. Although this flux is not able by itself to account for the ionization of the WIM, they suggest that it may be comparable or even dominant to the Lyman continuum flux escaping from the disk at large distances from the plane. In addition to contributions from multiple ionization sources, substantial evidence has been provided recently that the heating of the WIM is likely dominated at low densities by a source other than photoionization \citep{HRT99,RHT99}.

The development of high quantum efficiency, low read noise CCDs has recently allowed breakthrough \ha\ studies of the WIM. In addition to the velocity-resolved, all-sky survey described here, three large-area imaging surveys are also available or in progress. A southern survey with arc-minute resolution has been completed by \cite{SHASSA}. \cite{VTSLS} have been imaging the northern sky with similar resolution. A high-resolution ($\sim 1\arcsec$) survey of the Galactic plane and Magellanic clouds is also underway by \cite{UKST}. Furthermore, direct measurement of extremely faint, optical diagnostic lines in the WIM, including \hei\ $\lambda$5876 \citep{TuftePhD}, [\oi] $\lambda$6300 \citep{Reynolds+98,Hausen+02-aj}, and [\nii] $\lambda$5755 \citep{Reynolds+01}, have placed important constraints on the nature and source of the ionization and heating of this gas in our Galaxy.

Some of the first, most comprehensive work on the WIM (also referred to as Diffuse Ionized Gas---DIG---in this context) of an edge-on spiral galaxy similar to the Milky Way was undertaken by \citet{Rand97, Rand98} who measured numerous faint ionized emission lines far into the halo of NGC 891. \cite{DD97} and \cite{GDD96} also probed similar lines in the edge-on galaxies NGC 2188 and NGC 4631. All found a pronounced increase in the intensity ratios [\sii]/\ha\ and [\nii]/\ha\ in regions with fainter \ha\ emission. For spectra obtained perpendicular to the galaxies, these ratios rise systematically with distance from the galactic plane. More recent spectroscopic work has extended the detection of the WIM and these line ratios to impressive distances above galactic disks \citep{TD00,Rand00,CR01,Otte+01,OGR02,HW03}. Meanwhile, deep imaging of the extent and morphology of the WIM in spiral galaxies is providing valuable information about the link between the ionized layer and the star formation within galaxies \citep{HWR99,RD00,ZRB00,RossaPhD,Thilker+02,Zurita+02}.

Utilizing this new CCD technology, we designed and built a dedicated survey instrument, the Wisconsin H-Alpha Mapper (WHAM), which has produced the first deep, velocity-resolved maps of the WIM in our Galaxy as well as the first clear detections of \ha\ emission from high- and intermediate-velocity clouds \citep{TRH98, HRT99} and detections of extremely weak diagnostic lines from the disk and halo gas of the Milky Way \citep{TuftePhD,Reynolds+98,Reynolds+01}. The WHAM Northern Sky Survey (WHAM-NSS), the subject of this work, provides the first \ha\ survey of the distribution and kinematics of the diffuse \hii\ comparable to earlier 21 cm surveys of the \hi.

\section{OBSERVATIONS}

The WHAM instrument was installed at Kitt Peak in late 1996 and began obtaining survey data in early 1997. Much of the survey was complete after this first year. Filling the remaining holes, obtaining backup data for regions that were taken on nights of questionable quality, and extending the southern declination limit from $-20\arcdeg$ to $-30\arcdeg$ continued throughout 1998. Below, we describe the unique features of our instrument (\S\ref{sec:instrument}), the characteristics and strategies of the survey itself (\S\ref{sec:survey}), and the methods we use to transform the raw data into clean Galactic profiles (\S\ref{sec:data}).

\subsection{The WHAM Instrument}
\label{sec:instrument}

\begin{figure}
\plotone{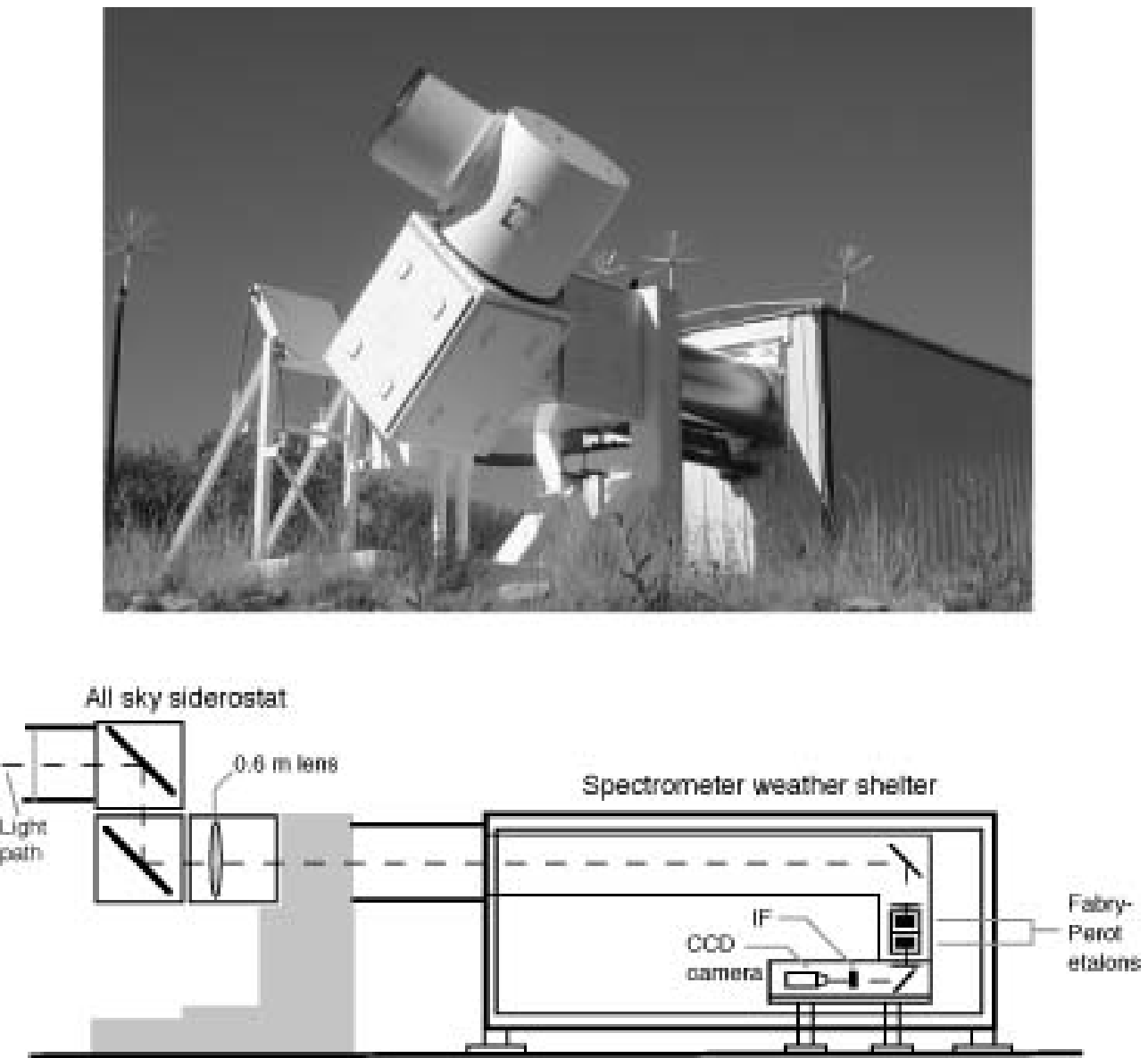}
\caption{The WHAM instrument. The top photo shows WHAM at Kitt Peak National Observatory, AZ. A schematic of the instrument is displayed in the bottom panel. The dashed line denotes the path of an on-axis ray through the system. As shown in the photo, both flat mirrors in the siderostat rotate independently, allowing all-sky pointing. \label{fig:wham}}
\end{figure}

\begin{figure}
\begin{center}\includegraphics[scale=0.75]{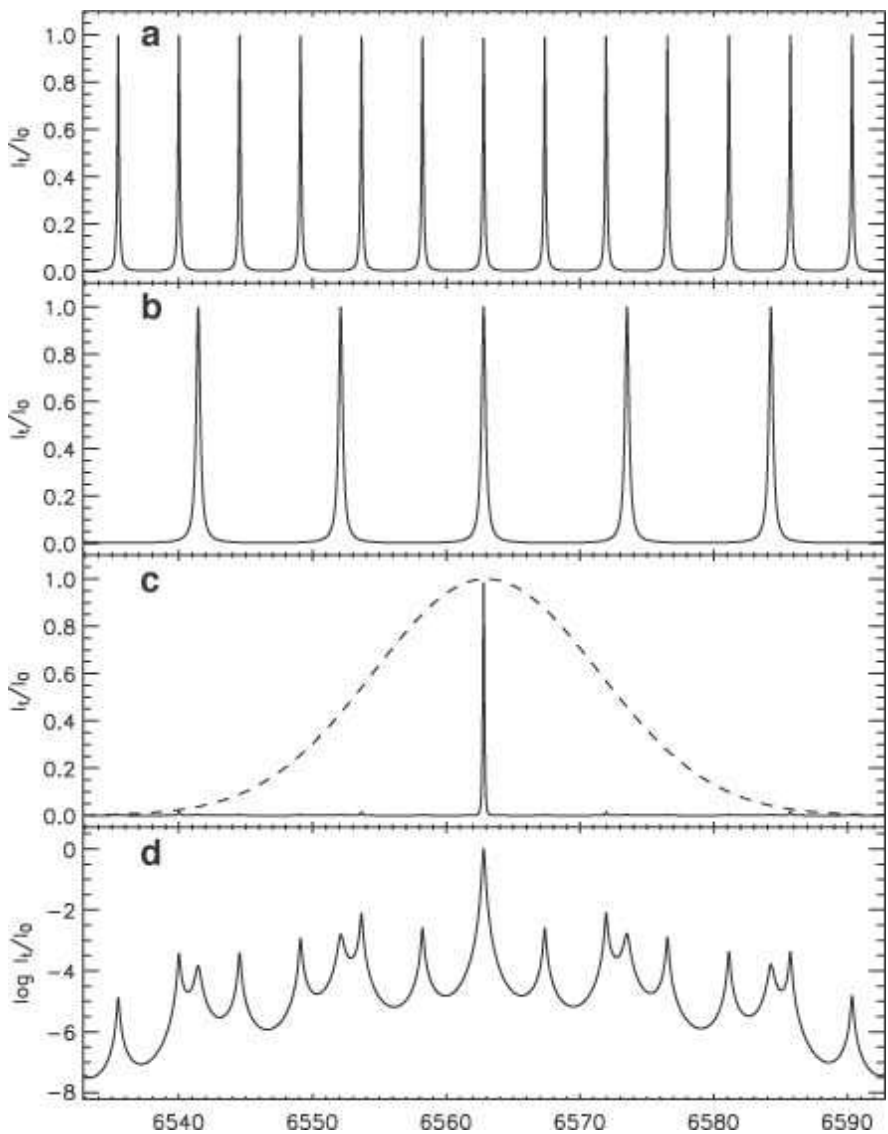}\end{center}
\caption{Etalon transmission functions. These curves depict the theoretical fraction of light transmitted through the \emph{(a)} high-resolution and \emph{(b)} low-resolution Fabry-Perot etalons used in WHAM. This example shows the transmission functions for a single incident angle with the etalons ``tuned'' (\S\ref{sec:instrument}) near \ha. When used in series, this pair of etalons  produces the composition transmission curve shown in \emph{(c)}. A theoretical filter profile with FWHM $\approx 20$ \AA\ is shown as the dashed line. \emph{(d)} plots the logarithm of the composite etalon and filter transmission function, showing the detailed structure of the residuals from the order suppression. \label{fig:dualetalon}}
\end{figure}

Due to the extended nature of diffuse ionized gas, large-aperture Fabry-Perot
detection techniques have proven to be quite successful in studying
the WIM \citep{RRSH90,Reynolds90}. WHAM continues to exploit this feature with its 15-cm aperture, dual-etalon Fabry-Perot spectrometer that delivers high-efficiency observations at high spectral resolution. WHAM is simply designed and consists of the following primary components, which are highlighted in Figure~\ref{fig:wham}:

\begin{itemize}

\item \textbf{Siderostat.} A custom-built all-sky siderostat houses two flat folding mirrors in a tipped ``alt-az'' mount. The system was designed to point and track continuously over the entire sky. The siderostat was designed and built by the University of Wisconsin--Madison Physical Sciences Laboratory.

\item \textbf{Pre-etalon Optics.} A pair of 860-cm focal length lenses provides telecentric illumination of the etalons. The stationary housing of the siderostat hosts a 63-cm diameter lens while a companion 18-cm lens sits in front of the first etalon surface. With this configuration, a one-degree beam on the sky is imaged into a 15-cm diameter region at the focal plane between the two etalons, and the 0\fdg50 half-angle cone from the sky is transformed into a 2\fdg09 half angle cone as it passes through the etalons.

\item \textbf{Etalons.} Two 15-cm etalons form the spectroscopic element for WHAM. The dual-etalon design combines a high-resolution etalon (with a gap spacing of 0.0471 cm) in series with a low-resolution etalon (with a gap spacing of 0.0201 cm). The low-resolution etalon suppresses higher orders from the high-resolution etalon, opening up a much larger useable spectral range in each observation than a single high-resolution etalon would provide, while suppressing the broad Lorentzian-like wings of the Fabry-Perot response profile. This feature is illustrated in Figure~\ref{fig:dualetalon}. With our chosen gap spacings, a reflective finesse of 30, and a maximum half angle of 2\fdg09, WHAM spectral observations deliver 12 \kms\ resolution (R $\approx 25,000$) over a 200 \kms\ spectral window near \ha. The location of the spectral window is selected by changing the pressure of a high index of refraction gas (SF$_6$) in sealed etalon chambers. This change in the index of refraction translates the transmission functions shown in Figure~\ref{fig:dualetalon} horizontally (i.e. in wavelength). One transmission peak from each etalon must also be coincident for maximum efficiency. We perform this alignment nightly as a part of our set-up procedure, and the pair of pressures that maximize the composite transmission function is called the ``tune'' for that spectral window. Since environmental effects can change the index of refraction as well, the tune can change slowly over day-long spans and must be confirmed every night.

\item \textbf{Post-etalon Optics.} To achieve even higher spectral isolation and to reject light outside the 4800 \AA\ to 8400 \AA\ band where the etalons are no longer reflective, the final rejection element is a standard 3.1'' (7.87 cm) narrow-band ($\sim 20$ \AA\ FWHM) interference filter with a transmission waveband centered near the emission line of interest (see Figure~\ref{fig:dualetalon}). WHAM is fitted with an eight-position filter wheel to allow observations of many emission lines during a night. A pair of 104-cm focal length lenses provides telecentric imaging of the Fabry-Perot ring pattern that emerges from the last etalon surface onto the interference filter. Finally, a four-lens stack (effectively f/0.9) images this ring spectrum onto a $\approx$ 1 cm diameter area of the CCD chip, while minimizing aberrations to less than 100$\mu$m (one $4\times4$ binned pixel).

\item \textbf{CCD.} WHAM's imaging device is a cryogenically cooled ($\approx -101\arcdeg$ C) Tektronix TK1024 with 24$\mu$m pixels and a quantum efficiency of 88\% at \ha. Our detector setup results in a gain of $0.57 e^{-}$/ADU and an effective read noise of $3.87 e^{-}$ rms after $4\times4$ on-chip binning.

\end{itemize}

Coatings on the etalons, optics, and CCD allow observations anywhere between 4800 \AA\ and 7320 \AA. All observation taken for the survey are centered around \ha\ and were obtained through a 20 \AA\ FWHM interference filter manufactured by Barr and Associates whose center wavelength is centered near 6565 \AA\ on axis, which translates to 6563 \AA\ (\ha) at the average angle of incidence onto the filter.

Note that in the configuration described above, WHAM images only the Fabry-Perot ring pattern emerging from the etalons. All spatial information is lost, with each point within the 1\arcdeg-diameter beam sampled equally by every spectral element within the 4.4 \AA\ (200 \kms) bandwidth of the spectrum. In this ``spectral mode,'' a CCD image contains the average spectrum within the one-degree diameter circular beam, avoiding any confusion between spectral features and angular structures of the source (or stars) within the beam. WHAM also has a set of optics that can be placed in the post-etalon beam to provide an extremely narrow-band (20--200 \kms), 3\arcmin\ angular resolution image of the sky within the one-degree field of view. However, only the spectral mode (with one-degree angular resolution) is used for the \ha\ Sky Survey observations presented here. Further details on the WHAM spectrometer and its calibration can be found in \citet{TuftePhD}.

WHAM was designed from its inception as a remotely operable facility. At the core of the operational software, WHAM implements the WIYN Telescope Control System \citep{Percival95}. All essential aspects of the instrument are able to be efficiently controlled by a computer over a low-bandwidth connection to the Internet. With this setup, WHAM observers have been able to maintain a demanding observing schedule without the associated costs in time and funding of long-distance travel.

\subsection{Survey Strategies}
\label{sec:survey}

\begin{figure}
\plotone{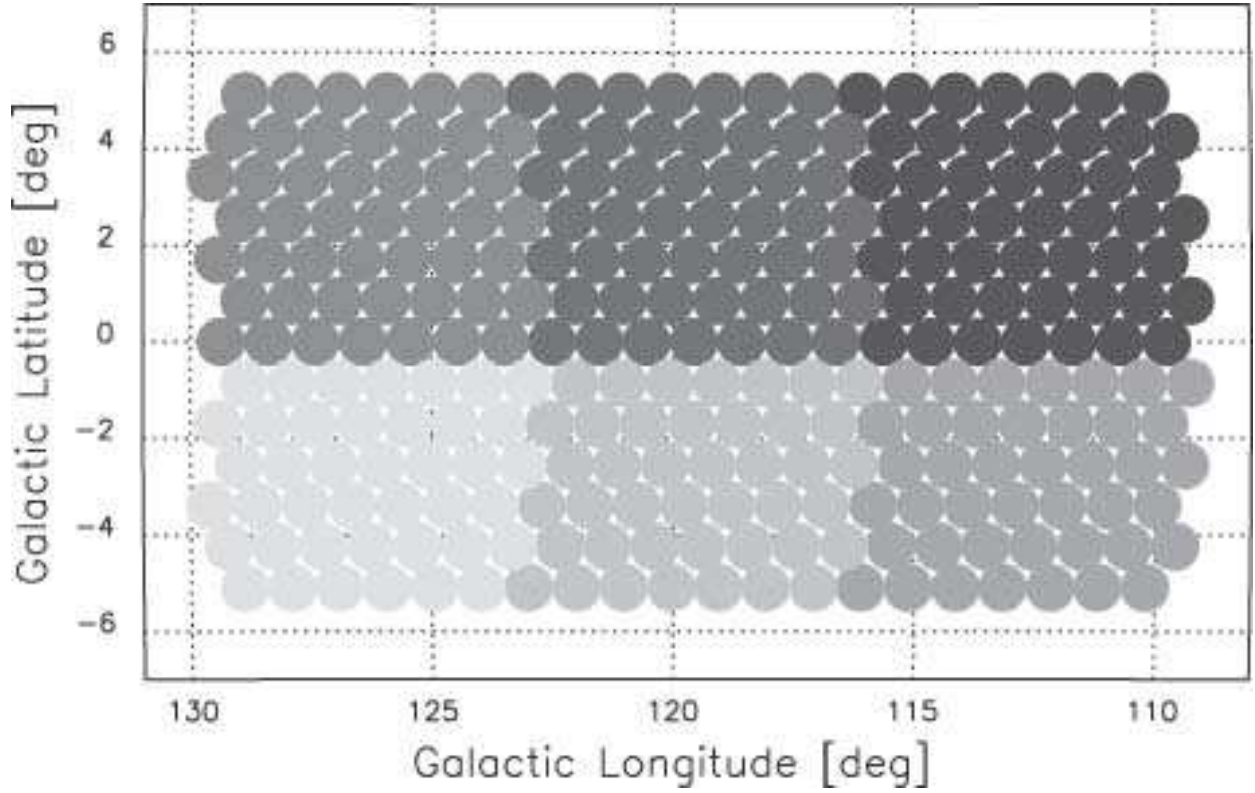}
\caption{Survey beam sample. Each circle represents a one-degree pointing of the WHAM survey. Pointings of the same shade of gray belong to the same observational ``block'' (see \S\ref{sec:survey}). \label{fig:beams}}
\end{figure}

The northern sky survey consists of more than 37,000 pointings of the one-degree beam on a grid with $\Delta b = 0\fdg85$, $\Delta l = 0\fdg98 / \cos b$, and $\delta \geq -30\arcdeg$. The origin of the longitude gridding is offset by ($0\fdg98/2) / \cos b$ between adjacent latitude rows. This odd spacing achieves nearly full sky coverage near the Galactic plane by nesting the beams. At higher latitudes, some ``slipping'' of the rows begins to occur and at some latitudes slightly larger gaps of unsampled sky are left between adjacent latitude rows. Since the goal was to probe the extended emission of the WIM, these small gaps have little effect on the completeness of the survey. To organize these observations, the individual pointings are grouped into observational ``blocks.'' Each block consists of up to 49 one-degree pointings, typically in a $7\times7$ nested grid. An example of our beam spacing and block division is presented in Figure~\ref{fig:beams}. Observations begin in one corner of a block and proceed in a boustrophedonic raster toward the opposite side, assuring that the sequence of observations within a block takes small steps in both time and on the sky. The exposure time for each one-degree pointing is 30 seconds, providing a sensitivity level that is better than 0.15~R (3$\sigma$). These approximately 30-minute blocks then form a tractable observational unit.

To minimize solar features in the background of our spectra, particularly the Fraunhofer \ha\ absorption line, survey observations were obtained only after astronomical twilight had ended and when the moon was below the horizon. In addition, all observations are made below a zenith distance, $z$, less than 60$\arcdeg$ (i.e. at an airmass, $Z$, less than 2) to avoid substantial atmospheric extinction corrections (see \S\ref{sec:ical}) and reduce the contamination of atmospheric lines (see \S\ref{sec:atmlines}). 

One final constraint attempts to maximize the spectral velocity separation between the geocoronal \ha\ emission line (the primary atmospheric contaminant in these spectra, see \S\ref{sec:geosub}) and the Local Standard of Rest (LSR). Due primarily to the orbital velocity of the earth and the peculiar velocity of the sun with respect to the LSR, this separation depends on both the direction in the sky and the time of the year. Since the bulk of the Galactic \ha\ emission occurs near 0 \kms\ LSR, we chose to observe blocks when the velocity separation between the LSR and geocentric frames was $\gtrsim 20$ \kms\ on a given night. With 12 \kms\ velocity resolution, these separations are sufficient to identify reliably the relatively bright but narrow geocoronal \ha\ emission line and subtract it from the spectrum. This criterion is satisfied for every direction on some night during the year except near the ecliptic poles, where the separation is limited to a maximum of 15 \kms. 

For most directions in the sky, there are two periods during the year that satisfy this constraint, one with the LSR blueward of geocentric zero as the earth's orbital vector points toward the general region of the look direction, and one a few months later with the LSR shifted to the red as the orbital vector points away from the look direction. Blocks observed with the former configuration (LSR at negative geocentric velocity) rise above the horizon later in the night, and we label these ``morning'' blocks. Those observed with the LSR at positive geocentric velocities are visible earlier and are labeled ``evening'' blocks. Although WHAM is continuously tunable and can locate the 200 \kms\ window anywhere within the filter bandpass, we chose to adopt two fixed tunes for these configurations, called the morning and evening tunes. Since most of our spectral contamination is atmospheric (and thus fixed with respect to the geocentric scale), this decision minimizes variations in the calibration phases of data reduction. One negative side effect is that the actual range of LSR velocities observed is not fixed from pointing to pointing. As a result, the spectral coverage of an individual observation ranges between LSR velocities of $[-138, +64]$ \kms\ and $[-71, +131]$ \kms, with most observations taken within $\pm10$ \kms\ of $[-116, +87]$ \kms. The entire WHAM-NSS does have complete spectral coverage between $-70$ and $+65$ \kms, however. 90\% of the survey pointings provide complete coverage from $-90$ to $+75$ \kms. 

\section{DATA REDUCTION}
\label{sec:data}

Due to the simplicity of the instrument and the relatively stable skies at Kitt Peak near \ha, reduction of our spectral data was straightforward. The creation of calibrated interstellar \ha\ spectra from raw WHAM images consisted of the following steps:

\subsection{Ring-Summing} 
\label{sec:ring}

\begin{figure}
\plotone{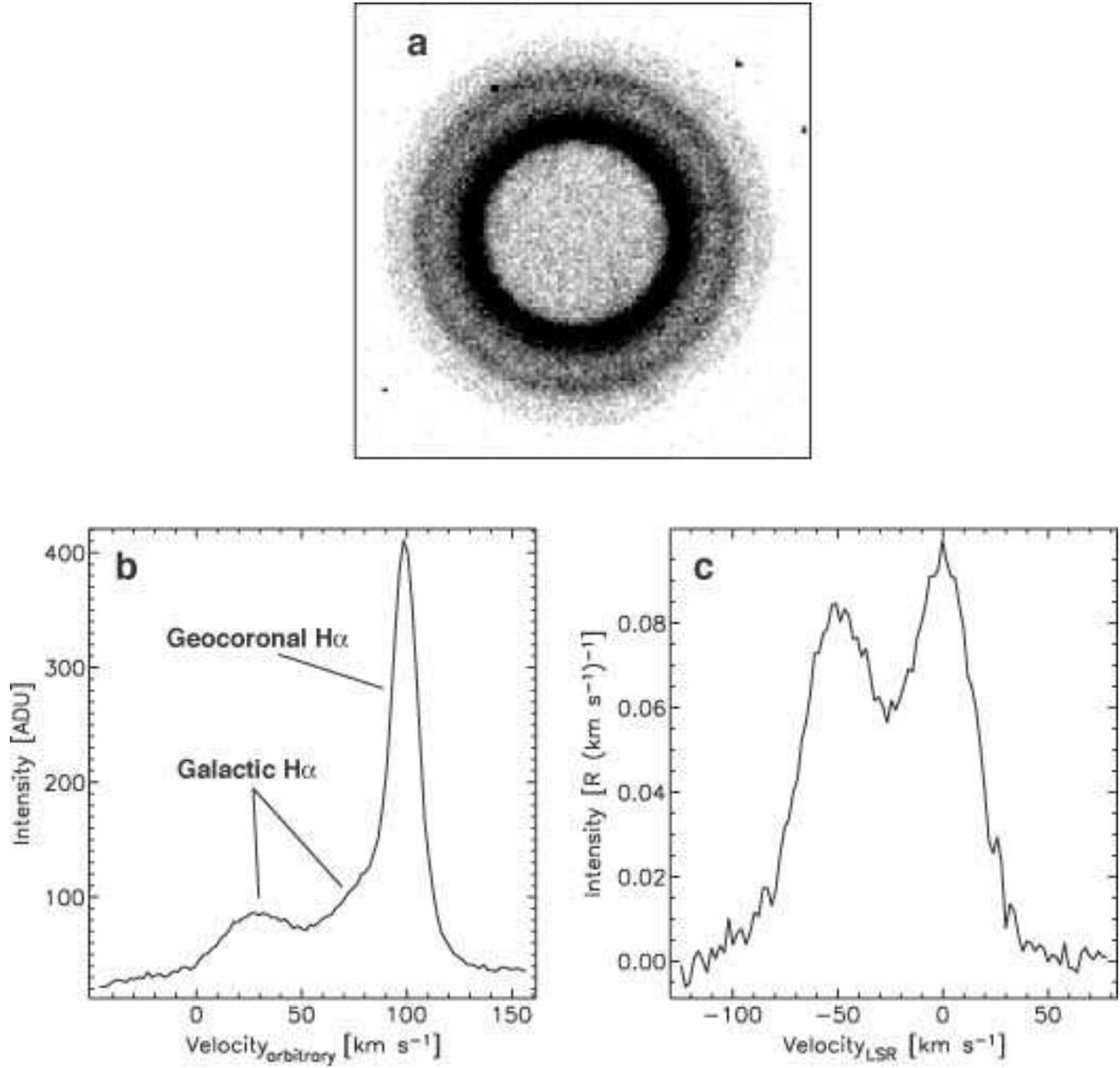}
\caption{WHAM data reduction. An unprocessed ring image is depicted in \emph{(a)}. This image is ``ring-summed'' (see \S\ref{sec:ring}) to produce \emph{(b)}. The fully-processed, calibrated spectrum in \emph{(c)} contains only Galactic \ha\ emission toward $\ell=124\fdg5$, $b=1\fdg7$.\label{fig:reduction}}
\end{figure}

Figure~\ref{fig:reduction}a shows the raw image output from the WHAM CCD camera. This ``ring image'' represents the spectrum within the one-degree beam. To reiterate, unlike many other astronomical Fabry-Perot instruments, the spectral mode of WHAM records \emph{only} spectral information---each part of the sky within the beam is sampled equally at all wavelengths and therefore no spatial information is obtained. Rings in these images are emission lines and the transition to zero signal at the largest 
radius represents the edge of our circular aperture defined by the 0.6-m diameter siderostat lens. Wavelength decreases radially from the center to the edge of the ring pattern, which is azimuthally symmetric. Since we are imaging only small angles ($\theta \lesssim 2\fdg09$) emerging from the etalons, $\Delta\lambda \sim v \sim \theta^2 \sim r^2$, where $r$ is the distance of a pixel from the center of the ring image \citep{TuftePhD}. Thus, equal-area annuli correspond to equal spectral intervals.

The image is converted into a one-dimensional spectrum by averaging the signal recorded in successive annuli (``annular-summing''; \citealp{Coakley+96,Nossal+97}). For these data, we define an 85-pixel (area) annular window for each data point and average the signal of all pixels whose centers fall within the radial limits for the annulus. This annular window size was chosen because it corresponds to a $\sim 2$ \kms\ spectral window, oversampling the width of our instrumental resolution by a factor of about six. The actual velocity width assigned to each window (equivalently an output spectral data point) has been determined empirically through observations of Hydrogen-Deuterium and Thorium-Argon calibration lamps. We measured the separation between several pairs of emission lines at different positions in the spectral field, allowing us to determine the first and second order coefficients of the velocity dispersion. The functional solution for our velocity calibration is then simply a second order polynomial:

\begin{equation}
v_{i}\ \mathrm{[\kms]}=v_{0}+2.083\;i_{r}-0.00036\;i_{r}^{2},
\end{equation}
 
where $v_{0}$ is an arbitrary zero point (at this stage, see \S\ref{sec:vcal} below for conversion to the LSR frame), $i_{r}$ is the integer index of a annular binning window, and $v_{i}$ is the velocity corresponding to that bin. $i_{r}=0$ corresponds to the outer-most (bluest) annulus.

Due to the discrete nature of the CCD pixels, these windows will not all have exactly 85 pixels, but the deviation in bin size is small so that the noise characteristics of the resulting data points are similar. We employ $4 \times 4$ on-chip binning to reduce read noise, but do not compromise spectral resolution at the blue end of the spectrum. The optical setup is designed so that the last 8 \kms\ spectral interval is one $4 \times 4$ ``super-pixel'' wide (96 $\mu$m on the chip) and fully sampled azimuthally.
  
Three other minor corrections also occur at this stage:

\begin{itemize}
    
\item \textbf{Bias subtraction.}
Bias frames are taken each night during observations. However, since we saw no significant structure in the bias images or drift in the average bias value during the first three years of operation, we reduced this step to subtracting a bias constant from each image.
    
\item \textbf{Cosmic ray removal.} 
Cosmic rays are removed in two steps. The first is a simple, $> 3\sigma$, neighbor-rejection technique that is used in many standard CCD image processing packages. We then take advantage of our azimuthal symmetry in the ring-summing to do a finer rejection. At a given radius, pixels are exposed to the same source intensity since they lie within the same spectral element. We fit a first-order polynomial to the $I$ vs.\ $r^2$ distribution within this annulus and apply another three-sigma rejection to pixels about this line. In a typical 30~s spectrum, less than 1\% of the image pixels are rejected with this multi-pass method.

\item \textbf{Reflection subtraction.}
A reflection with an intensity of  4\% that of the primary ring pattern is offset by $x = +37.50$, $y = -21.25$ pixels from the center of the image (very faintly visible in the lower right quadrant of Figure~\ref{fig:reduction}a). This feature arises from the reflection of the image off a wedged (0.01 rad), uncoated surface in the second etalon. To correct for this effect, the raw ring pattern is scaled, translated, and subtracted from itself before the ring-summing takes place. Other reflections in the system are at a significantly lower level and have no detectable affect on the resulting spectra.

\end{itemize}
  
Figure~\ref{fig:reduction}b shows the ring-summed spectrum from a typical WHAM observation.
  
\subsection{Flat Fielding} 
The majority of transmission inhomogeneities in the WHAM system can be removed by taking the spectral ring image produced by shining a diffuse, white light source through the entire optical path, processing it into a traditional spectrum as describe above, and dividing every spectrum by this resulting ``flat-field spectrum''. Although functionally quite different from dividing a spectral image by a white-light image and then ring-summing, we have found the former procedure to be much more robust for our application. Since roughly 85 pixels contribute to a single spectral data point, we find that the effect of intrinsic pixel to pixel gain variations is not significant for our data, especially with our second cosmic ray step. Pixels with significantly abnormal sensitivities are automatically rejected with this method, improving the estimate of the original source intensity.

This process effectively removes the primary vignetting shape produced by the geometry of the tandem Fabry-Perot etalons and the shift of the transmission function of the narrow-band filter with incident angle. Together, these factors result in a $\sim 20\%$ reduction in transmission from the center to the edge of the ring image.

\subsection{Extracting Galactic Emission}
\label{sec:extract}

Three components of our spectra, the background continuum, the geocoronal line, and fainter atmospheric lines, are removed in a single fitting and subtraction procedure. They are each described in more detail below. Their characterization and subtractions represents the bulk of the work to isolate pure Galactic emission in WHAM observations. Custom fitting software was designed to be semi-automated during the reduction of an observational block so that these corrections would be tractable with our large number of spectra. As noted in \S\ref{sec:survey}, pointings within a block are taken in small steps ($\sim 1\arcdeg$) across the sky and in time ($\sim 30$ seconds) so that reducing many of the blocks usually only requires hand-fitting the first pointing of a block. This solution is used as the initial input to the fit for the next pointing and the fitting steps along through the block. Galactic emission is modeled as a series of Gaussian components for this procedure, but we did not focus on the exact component structure of the emission in reducing the data. Instead we simply required a good fit to the underlying Galactic emission, especially in regions relatively free from atmospheric contamination. 

\subsubsection{Background Subtraction}

The brightness of the sky continuum emission at Kitt Peak within our one-degree beam at \ha\  is typically 15--30 ADUs, corresponding to approximately 0.022--0.044 R (\kms)$^{-1}$ (or 1.0--2.0 R \AA$^{-1}$), and is extremely flat within our 200 \kms\ bandpass (1 R $= 10^{6}/4\pi$ photons cm$^{-2}$ s$^{-1}$ ster$^{-1} = 2.4 \times 10^{-7}$ ergs cm$^{-2}$ s$^{-1}$ ster$^{-1}$ at \ha). Stellar sources in the one-degree beam with $m_{V} \lesssim 5$ also noticeably elevate the background level in the spectrum. A zero- or first-order polynomial is used in the fitting procedure to remove the background.  

If a bright star also has prominent absorption features in the spectral window (typically \ha), it can severely contaminate the spectrum and the blending of these stellar features with the emission makes recovery of the Galactic emission nearly impossible. In the publicly released data set, pointings that contain stars with $m_{V} < 5.5$ ($\sim 10\%$) are flagged to denote possible contamination. In the maps presented here, the relevant intensity integration of the pointing has been replaced by an average from neighboring spectra. WHAM may re-observe these directions with a small occulting disk near the imaging focal plane to eliminate the stellar contamination, but such observations are not included here.

\subsubsection{Geocoronal Subtraction}
\label{sec:geosub}

The primary contamination in the survey data arises from the geocoronal \ha\ emission line (see Figure~\ref{fig:reduction}b). This emission is produced primarily from excitation of neutral hydrogen in the outer layer of the earth's atmosphere through the scattering of solar Ly$\beta$ radiation ($n = 1 \rightarrow 3$). The subsequent decay cascade produces a prominent \ha\ emission feature in all spectra with an intensity of 2--13 R \citep{Nossal+01}. Since the intensity of the exciting Ly$\beta$ radiation depends most sensitively on the height of the earth's shadow along the look direction, the resulting strength of the contaminating \ha\ line depends on both the direction and time of the observation. For example, at a constant hour angle and declination, the intensity of this line can change by a factor of three from a minimum at midnight to a maximum at twilight.  At our spectral resolution, we do not resolve the fine structure of this line \citep{NRC98} and can fit it well with a single Gaussian component with an intrinsic width of 5--8 \kms. 

\subsubsection{Fainter Atmospheric Line Subtraction}
\label{sec:atmlines}

In addition to the geocorona, several much fainter ($< 0.2$ R), atmospheric lines of unknown origin are present near \ha\ in our spectra. As expected, the intensity of such lines rises with increasing zenith distance (i.e. airmass). The lines appear to be fairly stable over at least half-hour periods, but variations from night to night can be as large as a factor of two. These lines have been fully characterized by \citet{Hausen+02-apj}, and their measured parameters are reproduced here in Table~\ref{tab:atmlines}. 

To remove the atmospheric emission from the survey spectrum, we created a synthetic template of the lines using the positions, widths, and relative intensities listed in Table~\ref{tab:atmlines}. We hoped that this template could then be scaled by one parameter to fix the absolute intensity of all these lines together. However, an additional parameter was needed to adjust the intensity of line 10 at $v_\mathrm{geo}=+73$ \kms\ independently from the rest of the template. This ``line'' is affected by at least two known complications, which likely dictate the need for a second parameter in our template. First, the bright OH emission line located at 6553.6\AA\ is not fully rejected by our etalon configuration and a faint ``ghost'' line appears at this position near \ha. There is no a priori reason to suspect that the intensity of this ghost line is linked to the fainter lines seen near \ha. Second, there is a known absorption feature (probably H$_{2}$O) present at $v_\mathrm{geo} = +65$ \kms. The strength of this line is dependent on the brightness of the sky background on a given night toward a given direction, and will also depend on the column of H$_{2}$O along the line of sight. Its presence reduces the contribution of the other lines near these wavelengths, most substantially, the line at $v_\mathrm{geo} = +75$ \kms. In fact, we find a good correspondence between low values of the atmospheric parameter fit to the $v_\mathrm{geo} = +75$ \kms\ component and higher temperatures outside (when the water vapor column in the air can be higher). 

With this template in hand, we created very high signal-to-noise ``block-averaged'' spectra for each survey block (see \S\ref{sec:survey}) and fitted all the features (Galactic, geocorona, and atmospheric) in these spectra to empirically derive the two scaling parameters for the atmospheric template. Each of these block-averaged spectra represents a spatial average of about 30--50 square degrees on the sky and a temporal average over about 30 minutes. As noted above, the intensity of these faint atmospheric lines changes slowly over hour-long periods and in practice, we find that this method removes the contamination from each individual survey spectrum quite well. The template of weak atmospheric lines, scaled by these determined parameters, is used when determining the geocoronal fit (\S\ref{sec:geosub}) and is removed from the Galactic emission profile along with the geocoronal Gaussian. 
  
\subsection{Velocity Registration}
\label{sec:vcal}

As compensation for the contamination by the geocoronal line, we gain an accurate zero-point wavelength calibration in each \ha\ spectrum. Since the geocoronal line arises primarily from ground state excitation to $n = 3$ and not from recombination, the full assembly of fine structure \ha\ lines are not present and the line center of the geocoronal \ha\ multiplet is shifted -2.3 \kms\ (0.05 \AA) from the rest wavelength of the recombination line at 6562.82 \AA. Because we carry out a fitting of this line above (\S\ref{sec:geosub}), we use the record of the line center to register the geocentric velocity frame with an accuracy of less than 0.5 \kms\ in each spectrum. Translation from the geocentric frame to the Local Standard of Rest (LSR) is a simple calculation that depends on the direction and time of the observation. We use the same value for the Standard Solar Motion as that adopted by 21 cm observers, $+20$ \kms\ toward $\alpha_{1900} = 18\fh0$, $\delta_{1900} = +30\fdg0$.

\subsection{Intensity Calibration}
\label{sec:ical}

We observed ``standard'' nebular sources at least every 90 minutes during survey observations. All calibrations are tied to an absolute intensity measurement of the North American Nebula (NAN; NGC 7000) within a 50$\arcmin$ beam at 850~R by \cite{Scherb81}. Observations of NAN with WHAM yield the conversion: 1~R (\kms)$^{-1} = 684$ ADUs for a 30~s exposure, for an estimated 800~R within our one-degree beam centered at $\ell=85\fdg6, b=-0\fdg72$. In the absence of extinction, 1 R corresponds to an emission measure (EM$=\int n_{e}^{2}\,dl$) of 2.25 cm$^{-6}$ pc for gas at a temperature of 8,000 K, a typical value for the WIM \citep{HRT98, HRT99}.

By analyzing the nebular observations taken throughout each night at a variety of airmasses, we are able to compute the zenith transmission of the atmosphere at \ha\ for most nights of the survey. For nights where insufficient nebular observations exist for a specific determination, the atmospheric transmission is interpolated from an empirical fit to nights with well-determined transmission characteristics. We find that over the duration of the survey, the zenith transmission is described well by:
\begin{equation}
  T=T_{0}+A \cos(\frac{2\pi}{365} (d - \phi)),
\end{equation}
where $d$ is the day of the year, $\phi$ and $A$ are the phase and amplitude of the seasonal transmission variance, and $T_{0}$ is the yearly average transmission. The values for these parameters derived from our calibration observations are $T_{0}=0.924$, $A=0.026$, and $\phi=0.12$. However, the root-mean-squared scatter of the real observations about this fit is large, 0.06, due to daily variations. With a transmission estimate available for each night of data, all intensities have been corrected to their intrinsic values above the atmosphere with the standard factor, $T^{-1/\cos z}$, where $T$ is the zenith transmission of the given night and $z$ is the zenith distance of an observation. Using $T_{0}$, the average correction increases from 8\% near the zenith to 17\% near $z=60\arcdeg$.

Two other small systematic intensity corrections have been applied. The first accounts for a small systematic distribution in the intensities of fully atmospheric-corrected nebular observations that appears to track the difference in tune pressures used for a set of observations. We believe this arose due to the slight difference in tuning the elaton system when using a narrow source (the geocoronal \ha\ line) versus a broader-line source (diffuse nebulae). This correction can increase or decrease observed intensities by up to 5\%. The last correction tracks the general degradation of the optical transmission of the system (typically due to dust on optical surfaces). Over the 24 months of the survey, this rate has been roughly constant at about 0.0103\% per day, resulting in a gradually increasing correction of up to about 7.3\% over the time the survey data were collected.

\cite{SHASSA} find that our calibrated data match their independent determination to better than 10\%. 

\subsection{Zodiacal Absorption Correction}
\label{sec:zodi}

\begin{figure}
\plotone{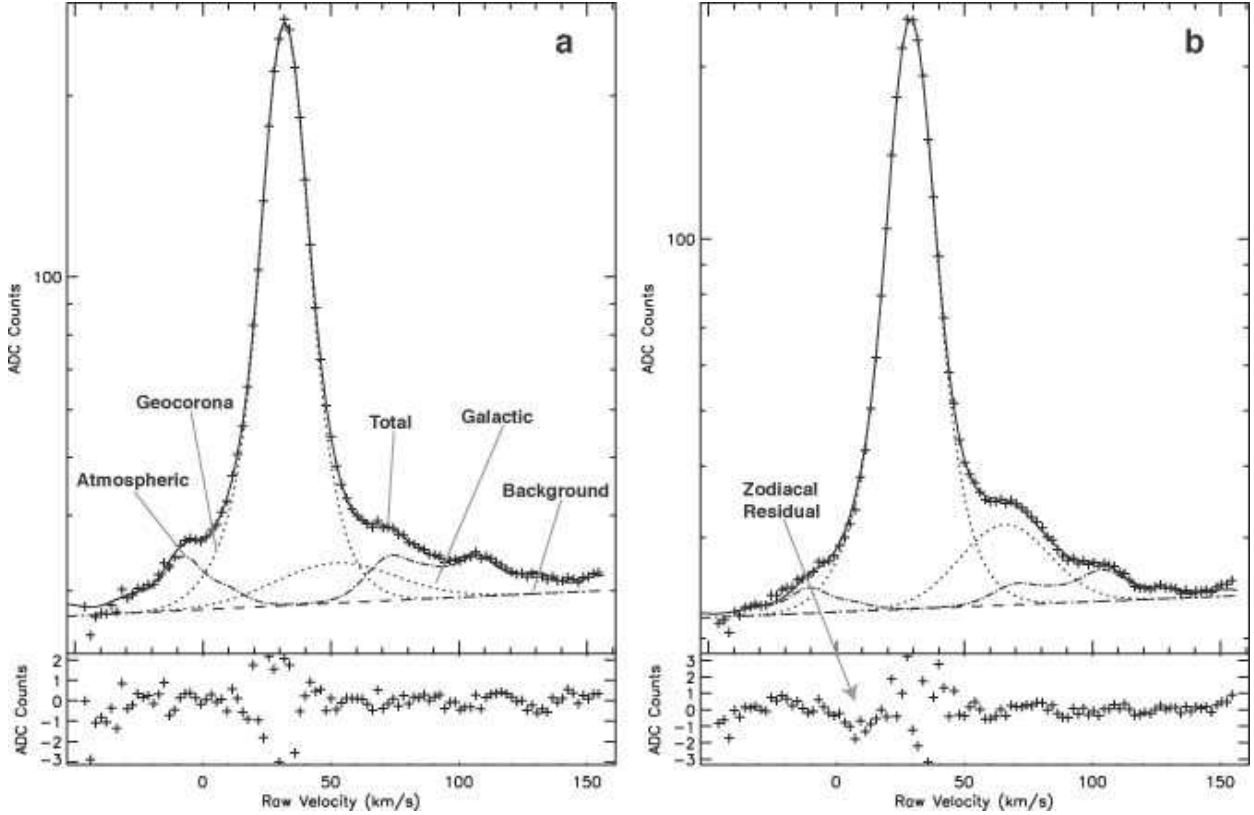}
\caption{Zodiacal Absorption Feature. Data and reduction fits are shown for two ``block-averaged'' spectra. Uncalibrated CCD counts are plotted logarithmically against a velocity frame with an arbitrary, but standardized zero-point. \emph{(a)} is the average spectra of the 41 pointings in the block centered near $\ell=167\fdg5, b=32\fdg2$. \emph{(b)} is the average spectra of the 42 pointings in the adjacent block centered near $\ell=174\fdg4, b=32\fdg2$. The data are plotted as the $+$ symbols. Each plot shows the total fit (solid line) as well as the individual components (labeled above) used to remove the various contaminating elements discussed in \S\ref{sec:data}. The residuals of each fit are plotted in the bottom panel of each figure. The increased scatter in the residuals around $v=30$ \kms\ is expected statistically due to the increased signal of the geocorona, whose peak contains more than 20 times more counts than other portions of the spectrum. \emph{(b)} reveals a systematic residual present in many blocks near the ecliptic plane. The feature is undetectable in \emph{(a)} because it is 5\arcdeg\ further from the ecliptic plane. \S\ref{sec:zodi} details our attempt to minimize the effect of this hidden feature, a portion of the zodiacal Fraunhofer absorption line. 
\label{fig:zodispectra}}
\end{figure}

As a testament to the sensitivity of WHAM, one subtle and unexpected feature remains in many of the survey spectra after all atmospheric emission lines are removed. Figure~\ref{fig:zodispectra} shows a pair of evening ``block-averaged'' spectra where the Galactic emission is simple (i.e. well-described by a single component) and faint ($\iha \approx 0.5$ R). A representative fit to each, including the background, Galactic, and atmospheric components (\S\ref{sec:extract}), is also shown, as well as the residuals from these fits. A subtle negative excursion remains in the residuals of Figure~\ref{fig:zodispectra}b slightly blueward of the geocoronal line. This residual feature is evident at some level in 10--20\% of the NSS blocks. Several pieces of evidence reveal that this feature is the solar Fraunhofer \ha\ \emph{absorption} line scattered off of zodiacal dust. Most importantly, those blocks observed with small angular distances to the Sun projected along the ecliptic plane (quantitatively known as the solar elongation, $\epsilon \equiv \lambda - \lambda_{\odot}$) and closest to the ecliptic plane delineate the anomalous features in our maps. Such close correlation with this coordinate frame points strongly to the feature arising from zodiacal light, whose intensity is also best described in this coordinate system. Also, the location of the feature relative to the geocorona (a fixed geocentric reference point near $v_\mathrm{geo}=0$ \kms) changes from evening to morning observations. In evening observations the feature appears blueward of the geocorona, while in the morning it appears redward. This change in the location of the feature from evening to morning is consistent with the expected velocity shift of the solar spectrum due to the orbital motion of the reflecting dust along these two different lines of sight where $\epsilon$ changes sign. The presence of this feature succinctly explains the deficiency of emission in evening spectra revealed by uncorrected channel maps centered at $\vlsr = -50$ \kms\ and in morning spectra by the $\vlsr = +50$ \kms. However, it also suggests that these contaminated spectra should be wholly corrected to recover emission removed by the presence of this broad absorption feature over the whole spectral window.  

Because the magnitude of this effect is small (typically $\lesssim 0.3$ R) and only a small portion of the absorption feature remains uncontaminated by emission features, we are unable to reconstruct the absorption profile from the data alone. Instead, we adopt a simple, global approach to minimizing this effect in the final data set. First, we assume that the zodiacal absorption line has the same shape as the solar Fraunhofer line. An electronic version of the spectrum of the solar \ha~line from \cite{Delbouille+73} is used to construct a template absorption profile. Second, we assume that the depth of the absorption line scales linearly with the visual surface brightness of the zodiacal light. Using the data provided in \cite{AllenBook}, we derive a smooth relationship for the V-band surface brightness as a function of $\epsilon$. Combining this function with a empirically determined constant scales the absorption profile to best fit our data. Finally, we use the velocity model of \cite{Fried78} to estimate the velocity shift we should apply to our model absorption line as a function of  $\epsilon$. We note that, prior to WHAM (Reynolds et al. 2003, in preparation), determinations of this final parameter have been fraught with large uncertainties, and studies have not agreed on the shape of the zodiacal rotation curve \citep{Fried78,ER84,Robley+85,LR+01}. However, given the confusion of the geocoronal emission and zodiacal absorption in the data under consideration, we believe that additional accuracy of these velocity shifts will not substantially aid the removal of this feature from the survey data.

Even with this model, accurately removing this feature from the spectra is not completely satisfactory. As Figure~\ref{fig:zodispectra} shows, the geocorona and Galactic emission obscure most evidence of an absorption feature. As a result, the strongest residual of the zodiacal feature in the most heavily contaminated spectra only occurs near one wing of the geocoronal line---that on the opposite side of the Galactic emission. With only this small area significantly constraining our overall scaling of the model, we find that a single factor is insufficient to adequately correct for the absorption line. The final scale factor for the evening spectra ($\epsilon < 0$) is 2.5 times greater than for morning spectra. The origin of this asymmetry is still unknown. The average strength of the line added back in to the spectra is $0.2$ R, with a maximum line strength of $0.4$ R for a few blocks taken particularly close to the Sun. Even with this correction applied to our final data set, special care should be taken when exploring survey data within 15 \kms\ of the geocentric rest wavelength of \ha\ because of the uncertainties associated with removing this zodiacal dust absorption feature. Of course, the removal of the much larger geocoronal emission feature also contributes to the uncertainty of the interstellar \ha\ spectrum near the earth's rest velocity.

\subsection{Maps}

Images (i.e. ``maps'') created from the survey data and presented here are constructed by first integrating each original spectrum over the velocity band of interest. Then, using the coordinates for the observations and these integrated intensities, our irregularly gridded data is translated onto a regular grid by Delaunay triangulation (IDL routines TRIANGULATE and TRIGRID). For the data presented here, observations contaminated by bright stars ($m_{V} \lesssim 5$) are excluded from the interpolation and replaced with an average of uncontaminated observations within one degree of the original pointing.

\section{SURVEY RESULTS}
\label{sec:results}

There are a plethora of new discoveries throughout the interstellar medium that are arising from this \ha\ survey and from other emission lines being probed by WHAM. Some of the early results have already been published elsewhere \citep{Reynolds+98, HRT98, TRH98, HRT99, RHT99, Lagache+00, Heiles+00, Reynolds+01, HRT01, RSH01, Hausen+02-apj, Tufte+02, Hausen+02-aj}. In this paper, we highlight the global results of the survey.

\subsection{Total Intensity Map}

\begin{figure}
\begin{center}\includegraphics[scale=0.75]{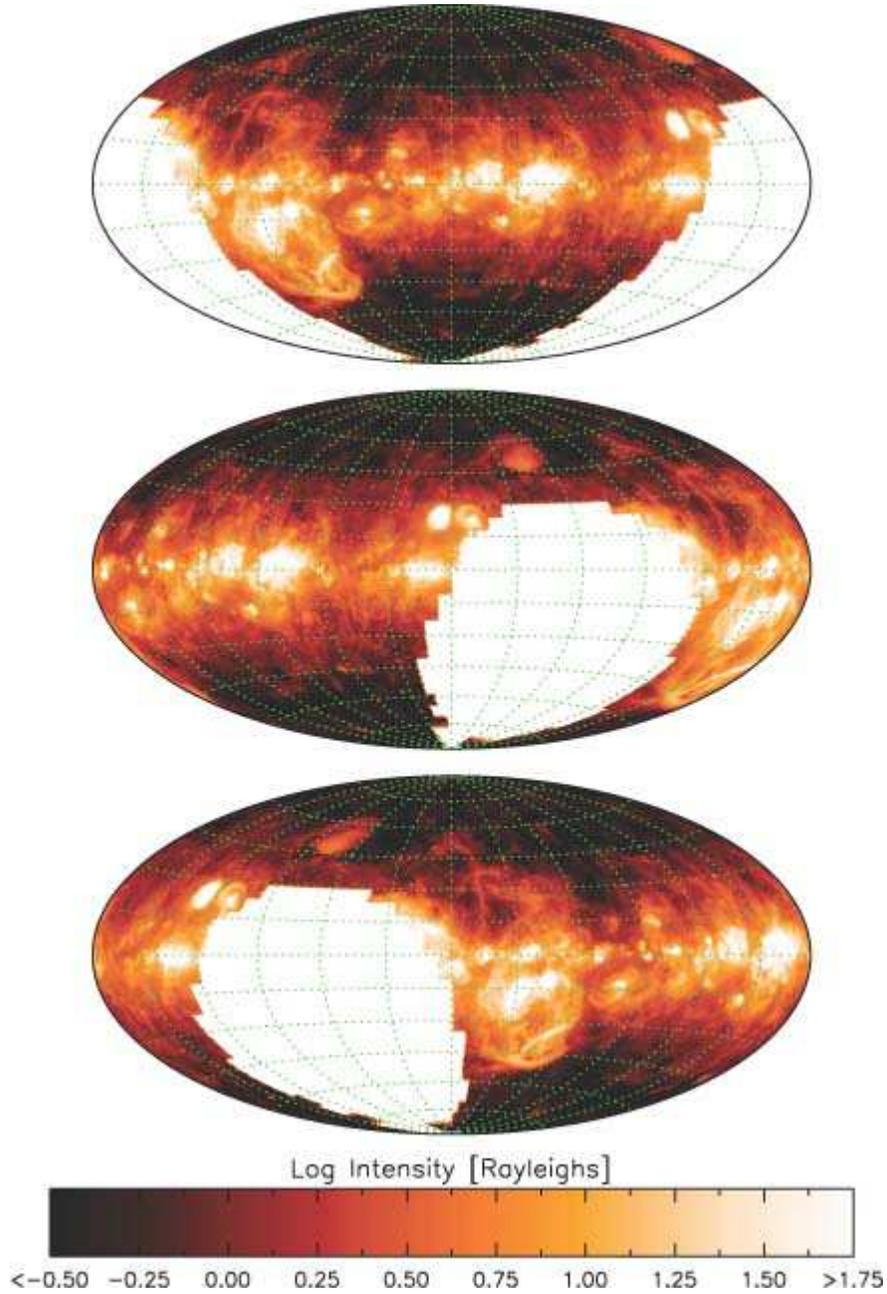}\end{center}
\caption{WHAM-NSS: Total Intensity. The integrated \ha\ intensity between $\vlsr=-80$ and $+80$ \kms\ is mapped as a function of Galactic coordinates. These three Hammer-Aitoff projections are centered at $b=0\arcdeg$ and $\ell=120\arcdeg$, 0\arcdeg, and 240\arcdeg, respectively from top to bottom. Dashed lines are spaced 30\arcdeg\ apart in longitude and 15\arcdeg\ apart in latitude. \label{fig:total}}
\end{figure}

Figure~\ref{fig:total} shows a series of three maps (centered at different projection centers) of the \ha\ emission from the northern sky, integrated between $\vlsr -80$ and $+80$ \kms. In addition to the bright knots of emission from discrete \hii\ regions in the plane, the general diffuse glow of the WIM is easily seen stretching to the Galactic poles. This material is rich in structure, with brighter filaments, loops, and regions superimposed on a fainter, smoother (or unresolved) background. As \S\ref{sec:quant} below discusses, the total emission is roughly described by a plane-parallel layer about the Galactic midplane. 

\begin{figure}
\plotone{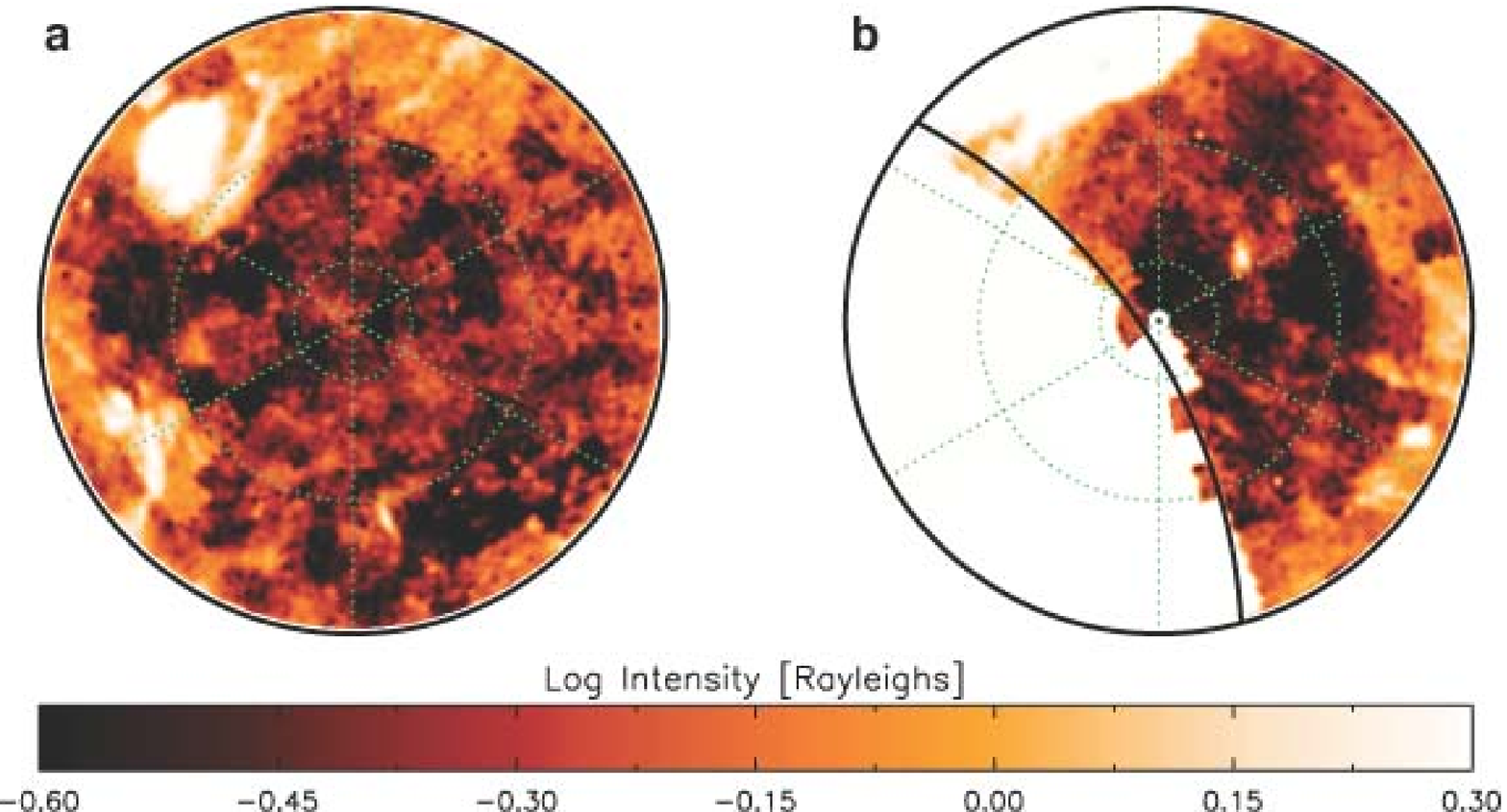}
\caption{WHAM-NSS: Total Intensity (Polar View). These stereographic projections map the total integrated \ha\ intensity between $\vlsr=-80$ and $+80$ \kms\ for the \emph{(a)} northern ($b=+40\arcdeg$ to $+90\arcdeg$) and \emph{(b)} southern ($b=-40\arcdeg$ to $-90\arcdeg$) Galactic polar caps. Green dashed lines are separated  by 60\arcdeg\ in longitude and 10\arcdeg\ in latitude. The grid line for $\ell=180\arcdeg$ is at the twelve-o'clock position. The solid line marking the border of data in the southern projection denotes $\delta=-30\arcdeg$, the limit of the WHAM-NSS. \label{fig:polar}}
\end{figure}

Due to the large dynamic range in the data, structure in the polar regions is lost in these all sky representations. Figure~\ref{fig:polar} focuses on only these regions, revealing that the diffuse \ha\ emission has detectable emission (and structure) everywhere in the sky. As more fully detailed in \S\ref{sec:quant} below, very few beams in the survey have total intensities below a few tenths of a Rayleigh. A few directions with very low (or no) emission detected in these short 30-second exposures have been studied in more depth by \citet{Hausen+02-apj}.

Table~\ref{table:total} gives a more quantitative representation of these integrated intensity Figures. 

\subsection{Channel Maps}

The power of the WHAM-NSS is revealed in Figure~\ref{fig:channels}. Here we present individual channel maps produced by using smaller integration ranges ($\Delta \vlsr = 20$ \kms). The Galactic plane now breaks up into emission produced by individual spiral arms. In particular, the outer Perseus Arm is prominent at negative velocities ($\vlsr < -30$ \kms) for $\ell = 90\arcdeg$ to $150\arcdeg$ and at positive velocities ($\vlsr > +30$ \kms) for $\ell = 210\arcdeg$ to $240\arcdeg$. The inner Sagittarius-Carina Arm is present primarily at ($\vlsr > +30$ \kms) for $\ell = 0\arcdeg$ to $50\arcdeg$. 

Many discrete features seen in \hi, especially those at intermediate velocities, are also seen in the WHAM-NSS (see \S\ref{sec:discussion}). A few prominent structures, such as the 50\arcdeg-long filament discussed in \citet{HRT98} rising near $\ell=225\arcdeg$ do not appear to have obvious neutral phase analogs. 

The spectral data from the WHAM-NSS is available through the Catalogues service at CDS (\url{http://cdsweb.u-strasbg.fr/}). 

\subsection{Longitude-Velocity Diagrams}

\begin{figure}
\begin{center}\includegraphics[scale=0.6]{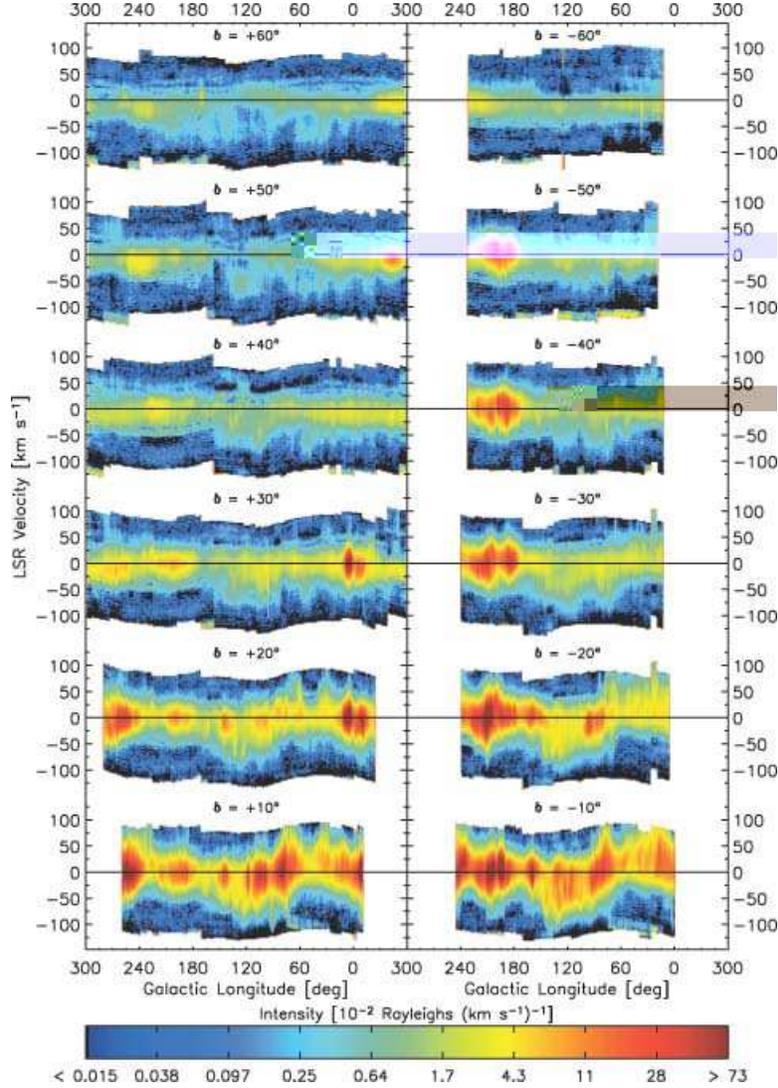}\end{center}
\caption{WHAM-NSS: Longitude-Velocity Diagrams. Each map displays the average longitude-velocity diagram for a 5-degree wide strip around the latitude listed above the map in steps of 10\arcdeg\ of latitude. Each map is centered around $\ell=120\arcdeg$. The left column plots positive latitudes ($b = +10\arcdeg$ through $+60\arcdeg$) and the right plots negative latitudes ($b = -10\arcdeg$ through $-60\arcdeg$). Both increase in distance from the Galactic plane toward the top of the figure. Since $b=0\arcdeg$ is heavily influenced by dust extinction, it is not shown (see text). The intensity of a point in the diagram is represented by logarithmically scaling the spectral values to a full-color table. For rough comparison to integrated intensities, a single component with FWHM $=30$ \kms\ and a peak value of $2.7\times10^{-2}$ R (\kms)$^{-1}$ has a total \ha\ intensity of 1 R.
\label{fig:lv}}
\end{figure}

We present a few representative $\ell$--$v$ diagrams in Figure~\ref{fig:lv}. Although they highlight the major features of the Galactic velocity distribution of the ionized gas including spiral arm delineation and velocity extent, they do not show detailed structure at the level found in displays of \hi\ and CO emission. This is primarily due to the lack of \ha\ components with narrow widths. The thermal broadening of the \ha\ line in gas with temperatures near $10^{4}$ K combined with any non-thermal motions result in line widths typically greater than 25 \kms. 

In addition, the effect of extinction by interstellar dust must be taken into account when using such diagrams to interpret global Galactic ionized gas distributions, especially within about 10\arcdeg\ of the Galactic plane. Above 10\arcdeg, these effects are small on average over the sky. However, very near the plane, some of the path lengths are known to be truncated by extinction, leading to potential confusion when interpreting a longitude-velocity diagram. We omit this particular latitude slice from Figure~\ref{fig:lv} as a result. 

\subsection{General Quantitative Analyses}
\label{sec:quant}

\begin{figure}
\begin{center}\includegraphics[scale=0.75]{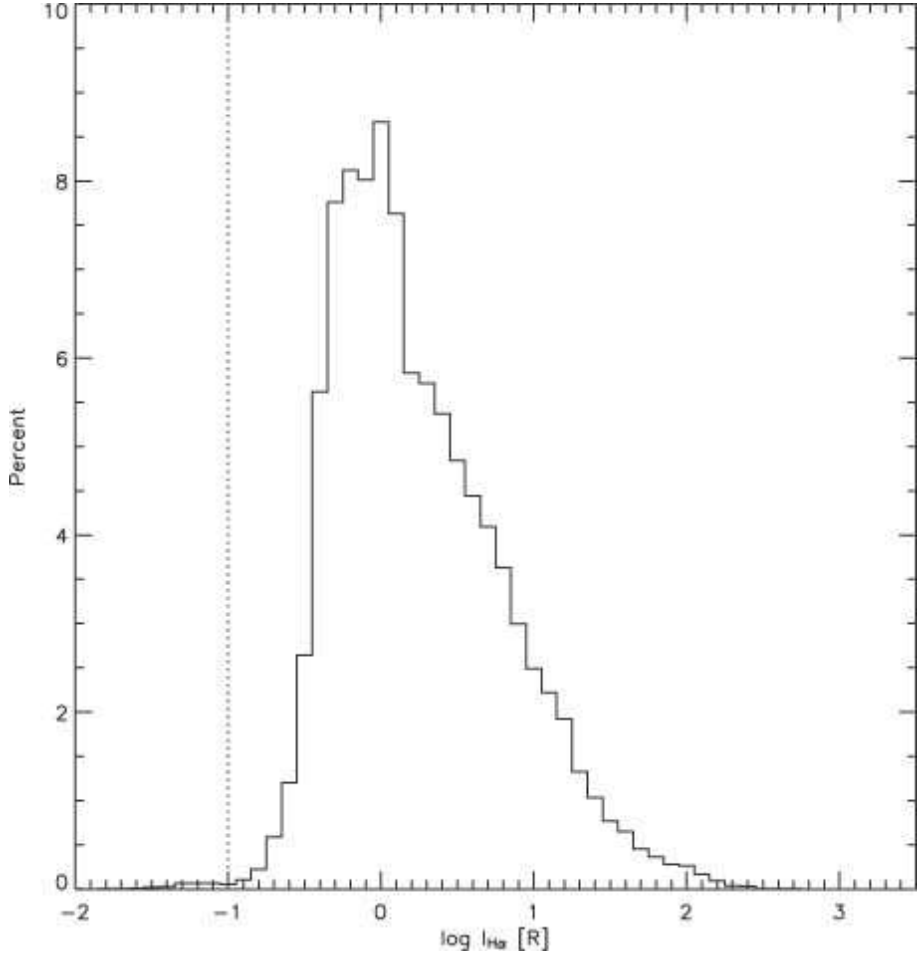}\end{center}
\caption{Distribution of Intensities. This histogram plots the distribution of the $\log \iha$ values in the WHAM-NSS. Notice that the steep drop-off near log \iha$=-0.5$ ($\sim 0.3$ R) is above the sensitivity of our instrument ($3\sigma \sim 0.15$ R, denoted by the dotted line). Within a one-degree beam, most directions have intensities within a decade of 1 R (EM $\sim 2$ pc cm$^{-6}$). \label{fig:histogram}}
\end{figure}

Inspired by the discussion of the global properties of \hi\ emission in \cite{DL90}, we present a similar analysis here for the \ha\ emission over the 3/4 of the sky covered by the WHAM-NSS. Figure~\ref{fig:histogram} presents the distribution of the \ha\ intensities as sampled by our one-degree beam over the sky. With the influence of discrete \hii\ regions, the \ha\ emission spans over five orders of magnitude even before any corrections are made for dust extinction and dilution by our one-degree beam. The few brightest pointings (in particular near the Orion Nebula, M42, and \hii\ regions in the plane toward the inner galaxy such as M20) measure over 1000 R. They are certainly not resolved by WHAM and have intrinsic \ha\ intensities much higher. On the faint end of the \ha\ intensity distribution, there is a very sharp cutoff between $\log \iha = -0.75$ and $-0.35$ ($0.2$ and $0.4$ R). This transition is comfortably above the sensitivity of the 30-second integrations used for the survey ($3\sigma \sim 0.15$ R). As with the neutral gas, the minimum emission in \ha\ is not at the Galactic pole, but in the direction of the Lockman Window near $\ell=152\arcdeg, b = +52\arcdeg$ \citep{JLM90,LJM86}. The regions of lowest ionized and neutral gas emission are not exactly spatially coincident but are within a few degrees of one another. A one-degree beam within the Lockman Window and a region of even lower \ha\ emission nearby have been examined in detail by \cite{Hausen+02-apj}.

\begin{figure}
\begin{center}\includegraphics[scale=0.75]{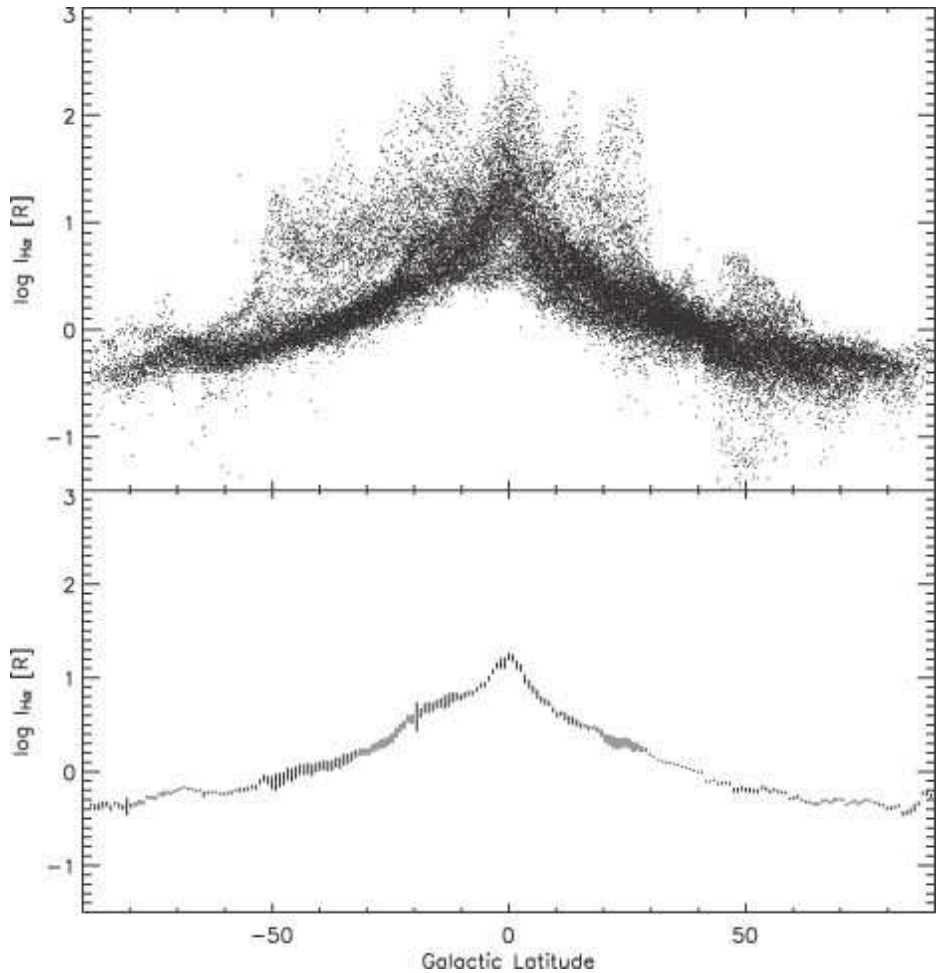}\end{center}
\caption{Intensity versus Galactic Latitude. The top panel plots the logarithm of the total intensity of every point in the WHAM-NSS as a function of Galactic latitude. Since the survey is regularly gridded in latitude, a small random perturbation (up to $\pm 0.85\arcdeg$) is applied to the true latitude of each point in the plot to avoid striation. The bottom panel plots the median value within each of the 0.85\arcdeg\ latitude strips from the survey. The vertical extent of each plotted bar is determined from the average deviation about the median within that bin. The larger deviations at $b=-10\arcdeg$ through $-50\arcdeg$ are due to the presence of the Orion-Eridanus superbubble (a brighter than average local feature) at these latitudes.\label{fig:i-vs-lat}}
\end{figure}

\begin{figure}
\begin{center}\includegraphics[scale=0.75]{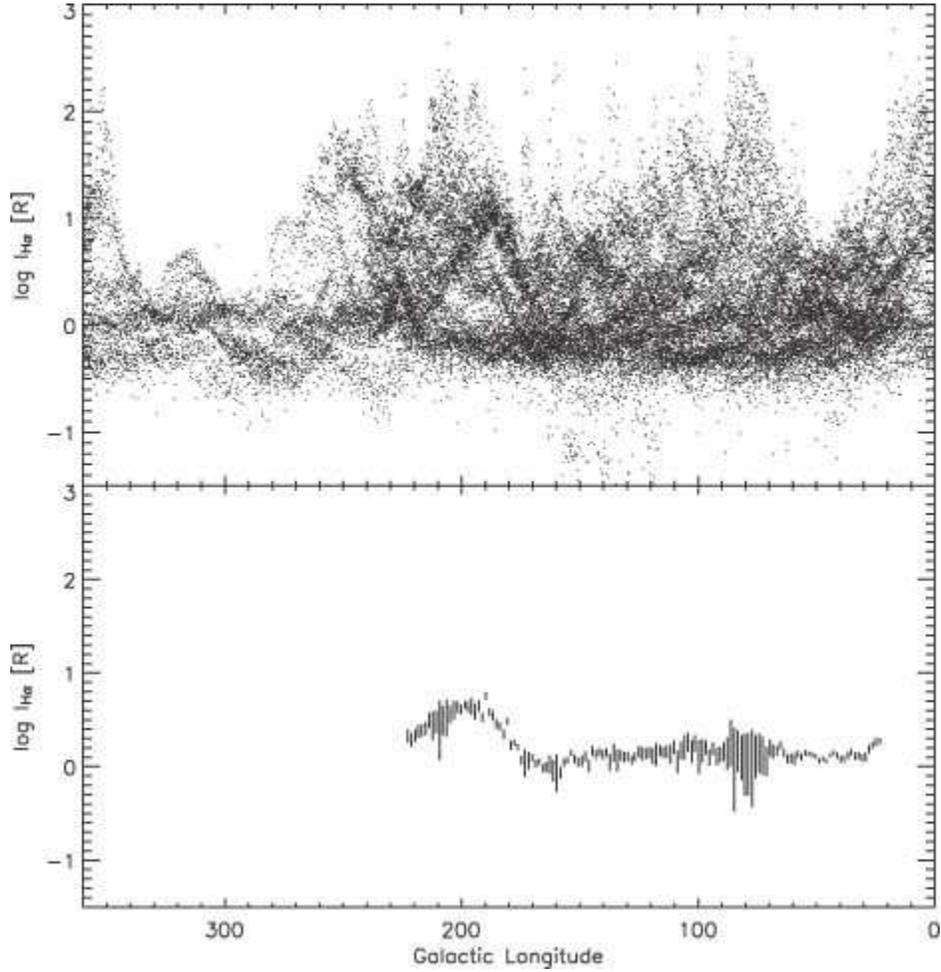}\end{center}
\caption{Intensity versus Galactic Longitude. The top panel plots the logarithm of the total intensity of every point in the WHAM-NSS as a function of Galactic longitude. The bottom panel plots the median value within one-degree longitude strips from the survey. The vertical extent of each plotted bar is determined from the average deviation about the median within that bin. These median values are only shown for the regions of the WHAM-NSS that are complete in latitude. \label{fig:i-vs-lon}}
\end{figure}

Figures~\ref{fig:i-vs-lat} and \ref{fig:i-vs-lon} examine the distribution of total \iha\ as a function of Galactic coordinates. The top panel of both figures plot each pointing in the survey while the bottom panel shows the median and spread (using the average deviation) within small bins ($\approx 1\arcdeg$) along the coordinate axis. Although using the median statistic does remove some of the influence of anomalously bright regions, the imprint of a few very large emission regions persists nonetheless. In particular, the Orion-Eridanus superbubble covering the large region from $\ell = 180\arcdeg$ through $220\arcdeg$ and $b = 0\arcdeg$ through $b = -50\arcdeg$ \citep{HHR99} causes a substantial enhancement in the trend of the median and size of the spread in both figures. 

\begin{figure}
\begin{center}\includegraphics[scale=0.75]{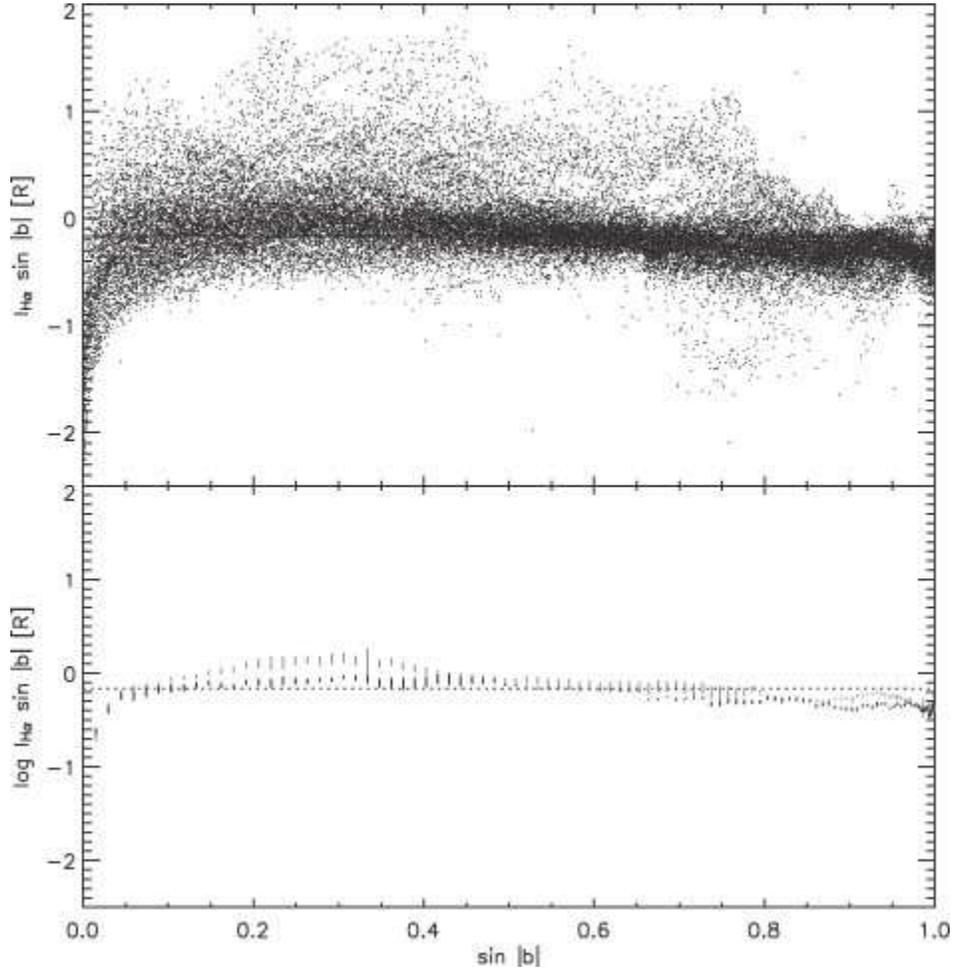}\end{center}
\caption{$\iha\sin |b|$ vs $\sin |b|$. If \ha\ emission were to arise from a plane-parallel slab, this representation would show intensity constant at all latitudes. The two tracks in the bottom panel divide the data into positive (black) and negative (gray) Galactic latitudes. Emission near $\sin |b|=0$ is likely influenced by extinction. The dotted line denotes the median $\iha \sin |b|$ value for the survey pointings. \label{fig:isinb-vs-sinb}}
\end{figure}

The sharp cutoff at low intensities in Figure~\ref{fig:histogram} and the distinctive shape of the latitude profile of the \ha\ emission in Figure~\ref{fig:i-vs-lat} can be explained to some extent by adopting a simple model of the ionized gas as a uniform plane-parallel layer. In this case, the minimum path through the layer at the Galactic poles provides a well-defined minimum intensity for any direction while the increase of the emission at low $b$ is mostly due to the geometrical effect of longer pathlengths through the layer. In such an idealized model, this cosecant effect is removed by plotting $\iha \sin |b|$ versus $\sin |b|$, as shown in Figure~\ref{fig:isinb-vs-sinb}. The top panel again plots every point in the WHAM-NSS and the bottom panel plots the median and average deviation, with the trend for the northern Galactic hemisphere ($b > 0\arcdeg$) split from that of the south ($b < 0\arcdeg$). This split is useful for two reasons. First, the longitude coverage of the north is much more complete. Second, the Orion-Eridanus superbubble (seen as the enhancement between $\sin |b| = 0.1$ and $0.4$) only contaminates the southern data.

\section{DISCUSSION}
\label{sec:discussion}

A full exploration of the rich detail in the WIM uncovered by the WHAM-NSS is beyond a single work. Here we comment on a few of the global observations highlighted by the presentation of the data in the various forms of \S\ref{sec:results}. 

\begin{figure}
\begin{center}\includegraphics[scale=0.8]{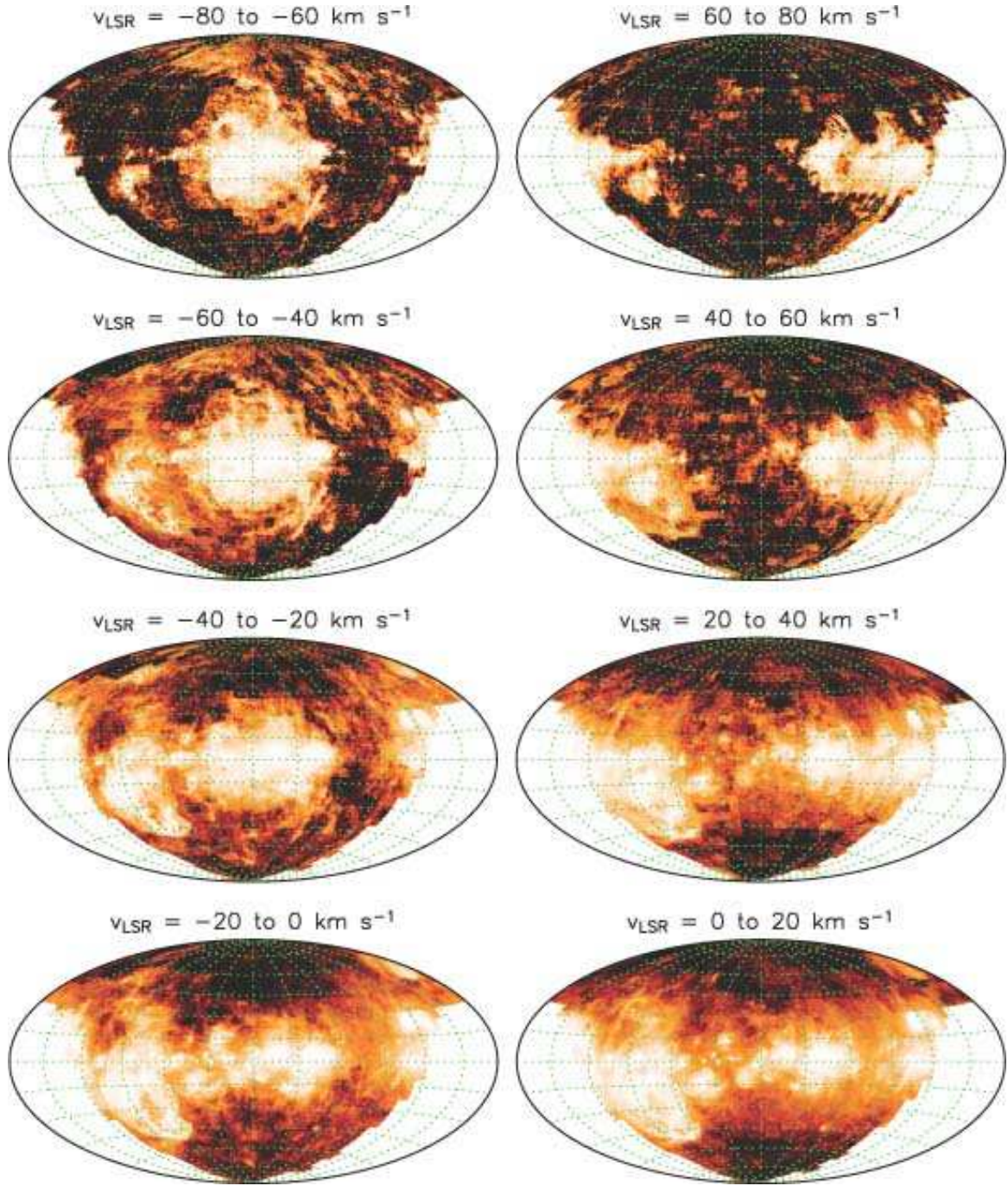}\end{center}
\caption{WHAM-NSS: Channel Maps. This series of maps shows the kinematic distribution of \ha\ emission in the WHAM-NSS. Each map is integrated over the 20 \kms\ range listed at the top of the map. Each map is scaled independently through histogram equalization to highlight the substructure within the channel. \label{fig:channels}}
\end{figure}

As also discovered by the recent deep \ha\ imaging surveys \citep{SHASSA, VTSLS}, the distribution of the ionized gas in the Milky Way is characterized by both a pervasive background of faint emission and rich filamentary structure. Now, with the absolute calibration provided by WHAM and Galactic emission profiles free from atmospheric and solar system contamination, we can quantify the background accurately and, in many cases, isolate the brighter features spectrally. \cite{HRT98} presented one of the more striking features evident in Figures~\ref{fig:total} and \ref{fig:channels}, a $> 50\arcdeg$ long filament stretching out of the Galactic plane near $\ell=225\arcdeg$, showing changes in its radial velocity along its length. Unlike most other distinct features in the ionized gas, this filament appears to have no counterpart in the neutral phase. \cite{RSH01} explore a large arc of emission above the Perseus arm near $\ell=135\arcdeg$ (most visible in the lowest two velocity channels of Figure~\ref{fig:channels}). When the NSS was completed, a similar arc was discovered south of the plane, directly opposite of the northern extension \citep{MHR02}. These massive structures were shaped out of the interstellar medium some time ago, but are now actively ionized by younger OB associations below them in the Galactic plane. Many of the intermediate-velocity complexes cataloged by \cite{Wakker01} also have associated \ha\ emission revealed in the WHAM-NSS. \cite{HRT99} discuss the character of one of the largest revealed in our survey, Complex K. The discovery of this associated ionized gas will not only aid the characterization of the intermediate- and high-velocity clouds \citep{Wakker+99}, but also provides one of the only direct probes of the ionizing flux in the halo \citep{TRH98,B-HM99,Tufte+02}. Finally, the survey has revealed several new \hii\ regions around early B and hot subdwarf stars \citep{Haffner01}. These discoveries could be important by helping to characterize a population of sources capable of providing ionizing flux away from the heavily absorptive Galactic disk. 

\begin{figure}
\begin{center}\includegraphics[scale=0.73]{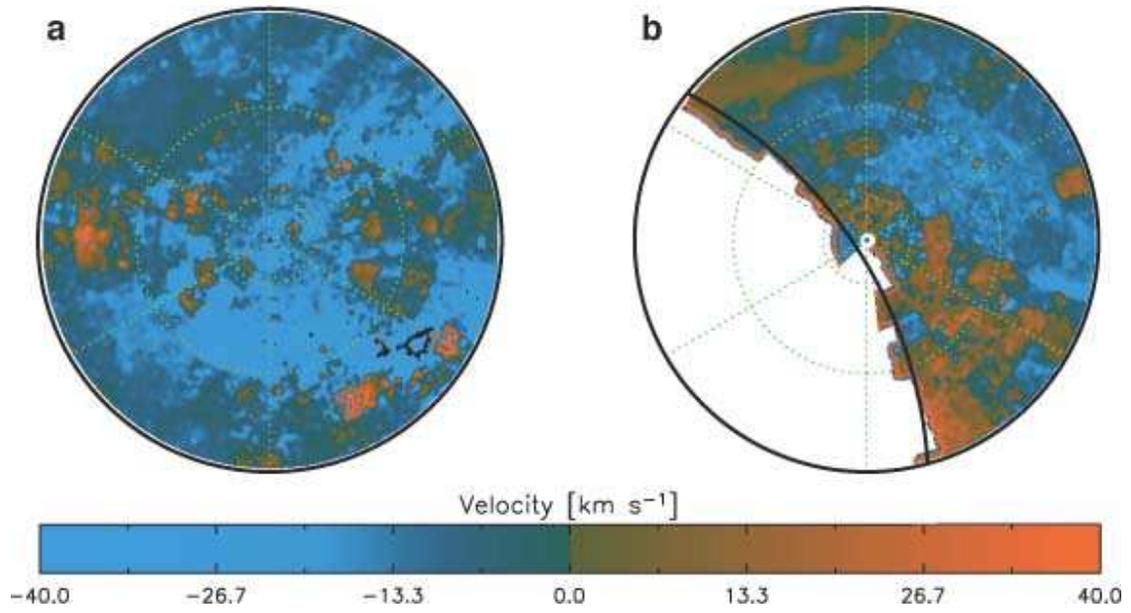}\end{center}
\caption{Velocities at high-latitudes. These maps mirror the projection aspects of Figure~\ref{fig:polar} but display intensity-weighted velocity averages (see \S\ref{sec:discussion}), which highlight a negative velocity preference for ionized emission toward both Galactic poles. \label{fig:polar-vel}}
\end{figure}

The kinematics of the WIM are broadly summarized by Figure~\ref{fig:lv}. Two general trends are worthy of special note. First, the kinematic imprint of spiral arms in both the inner and outer Galaxy is clearly seen even 30\arcdeg\ off the plane. Above this height, only a few discrete features that stretch to large heights above arms and intermediate-velocity complexes still show substantial non-local velocity patterns. The second general trend is that at high latitudes the \ha\ emitting gas is clearly biased toward negative velocities. The emission at $|b| = 60\arcdeg$ on both sides of the Galactic plane shows an asymmetric distribution about the solid line that denotes $\vlsr=0$. This trend is also seen in the neutral gas toward the Galactic poles \citep{Blitz+99}. Figure~\ref{fig:polar-vel} shows the spatial distribution of this effect by presenting maps of the average intensity-weighted velocity ($\int \iha v\,dv / \int \iha\,dv$) over both Galactic poles. In particular, it highlights the fact that deviations from zero velocity occur over coherent regions of rising and falling gas.

Even when the localized impact of bright \hii\ regions is minimized (\S\ref{sec:quant}), a plane-parallel model turns out to be only approximate in describing the global WIM, as is also the case for the neutral gas \citep{DL90}. The dotted line in Figure~\ref{fig:isinb-vs-sinb} delineates the overall median value of $\log \iha \sin |b|$ for the survey, $-0.17$, corresponding to an $\iha \sin |b| = 0.68$ R. Extinction by dust likely creates the observed decrease in $\iha \sin |b|$ at $\sin |b| < 0.05$ (i.e. $|b|<3\arcdeg$). Aside from this expected effect, instead of being constant, the Galactic emission decreases by a factor of two in the northern hemisphere from $\sin |b| = 0.2$ through $1.0$. Even with the limited longitude coverage in the south, the trend appears to be similar. As pointed out by \cite{DL90} to explain the similar effect in the \hi\ distribution, the presence of the Local Bubble \citep{Maiz-Apellaniz01,SC01,CR87}, a localized region of very low density ionized gas around the Sun, likely causes at least some of the decrease seen here. However, the much larger scale height of the \ha\ emitting gas \citep{HRT98,Reynolds97} should reduce the imprint of this local feature compared to its importance in the same analysis of the neutral gas. Two other effects unique to the ionized gas emission may also contribute to the slope seen here. As discussed in \cite{HRT98} and \cite{Reynolds97}, the vertical electron distribution derived from \ha\ observations appears to be fairly well described by a two-parameter exponential. However, these parameters may be determined primarily by local star formation so that, in particular, massive spiral arms may have distinctly different sets of parameters compared to the local neighborhood in which the Sun resides. In this scenario, low-$b$ sightlines will be preferentially elevated, resulting in a decrease in $\iha \sin |b|$ with increasing $\sin |b|$ as the local parameters become increasingly influential in the total emission along the line of sight. Finally, the increase at lower latitudes may be due in part to contamination by \ha\ scattered by interstellar dust that is far from the emitting source \citep{WR99}. 

\section{SUMMARY}

We have completed the first velocity-resolved survey of \ha\ emission over the northern sky. The WHAM-NSS sheds new light on the distribution and kinematics of the WIM, allowing comparative studies on a global scale with other phases of the interstellar medium. The survey reveals that:

\begin{itemize}

\item Ionized gas in the Milky Way is ubiquitous, with a pervasive yet faint background ($\iha > 0.1$ R) detectable in nearly every direction in the northern sky. 

\item The total \ha\ emission is roughly described by a plane-parallel layer about the Galactic midplane having $\iha \sin |b| = 0.68$ R, although there are noticeable systematic deviations from this average value.

\item The vertical distribution of the ionized gas is more extended than that of the neutral gas \citep{HRT98,Reynolds97}. 

\item The average kinematics of the ionized gas follow the velocity patterns of the nearby spiral arms and the imprint of the arms is able to be traced to a height of $|b| = 30\arcdeg$ in most cases.

\item At high latitudes, the kinematics of the ionized gas is preferentially biased toward negative velocities. 

\item Many of the anomalous intermediate-velocity features ($|\vlsr| < 100$ \kms) discovered in \hi\ surveys also contain detectable amounts of ionized gas \citep{HRT99}. WHAM's detection of \ha\ emission from high-velocity clouds \citep{TRH98} is not covered by this survey.

\end{itemize}

The WHAM-NSS along with the new deep \ha\ imaging surveys by \citep{SHASSA} and \citep{VTSLS} open a host of new opportunities to study the WIM. WHAM is continuing to explore the ionized gas of the Milky Way by extending observations to higher velocities and adding additional diagnostic emission lines in regions of interest. 

The full WHAM-NSS is available at \url{http://www.astro.wisc.edu/wham/}. 

\acknowledgments

The success of WHAM and its NSS is due to the contributions of many. In particular, we wish to thank Nikki Hausen, Brian Babler, Rebecca Pifer, and Mark Quigley, who helped with the monumental task of data reduction. The smooth remote operation of WHAM has been aided by the wonderful staff at Kitt Peak National Observatory. Trudy Tilleman's dedicated monitoring of the night sky conditions during the survey helped maintain confidence in the quality of our data. The siderostat was designed and built in collaboration with the UW--Madison Physical Sciences Laboratory, and the spectrometer was designed and built with the help of the UW--Madison Space Astronomy Laboratory. John Harlander provided crucial assistance with the optical design. The WHAM-NSS is the culmination of a decades-long program, initiated by Elihu Boldt and nurtured by Frank Scherb and Fred Roesler, to develop Fabry-Perot spectroscopy for the detection and study of diffuse interstellar ionized hydrogen. This research has made use of the SIMBAD database, operated at CDS, Strasbourg, France.

This work was supported by the National Science Foundation through grants AST 91-22701, AST 96-19424 and AST 02-04973, with assistance from the University of Wisconsin's Graduate School, Department of Astronomy, and Department of Physics.


\begin{thebibliography}{78}
\expandafter\ifx\csname natexlab\endcsname\relax\def\natexlab#1{#1}\fi

\bibitem[{Bennett {et~al.}(2003)Bennett, Hill, Hinshaw, Nolta, Odegard, Page,
  Spergel, Weiland, Wright, Halpern, Jarosik, Kogut, Limon, Meyer, Tucker, \&
  Wollack}]{WMAPFG}
Bennett, C.~L., Hill, R.~S., Hinshaw, G., Nolta, M.~R., Odegard, N., Page, L.,
  Spergel, D.~N., Weiland, J.~L., Wright, E.~L., Halpern, M., Jarosik, N.,
  Kogut, A., Limon, M., Meyer, S.~S., Tucker, G.~S., \& Wollack, E. 2003, \apj,
  submitted

\bibitem[{{Bland-Hawthorn} \& {Maloney}(1999)}]{B-HM99}
{Bland-Hawthorn}, J. \& {Maloney}, P.~R. 1999, \apjl, 510, L33

\bibitem[{{Blitz} {et~al.}(1999){Blitz}, {Spergel}, {Teuben}, {Hartmann}, \&
  {Burton}}]{Blitz+99}
{Blitz}, L., {Spergel}, D.~N., {Teuben}, P.~J., {Hartmann}, D., \& {Burton},
  W.~B. 1999, \apj, 514, 818

\bibitem[{{Coakley} {et~al.}(1996){Coakley}, {Roesler}, {Reynolds}, \&
  {Nossal}}]{Coakley+96}
{Coakley}, M.~M., {Roesler}, F.~L., {Reynolds}, R.~J., \& {Nossal}, S. 1996,
  \ao, 35, 6479

\bibitem[{{Collins} \& {Rand}(2001)}]{CR01}
{Collins}, J.~A. \& {Rand}, R.~J. 2001, \apj, 551, 57

\bibitem[{{Cox}(2000)}]{AllenBook}
{Cox}, A.~N. 2000, {Allen's astrophysical quantities} (New York: AIP Press
  (Springer-Verlag))

\bibitem[{{Cox} \& {Reynolds}(1987)}]{CR87}
{Cox}, D.~P. \& {Reynolds}, R.~J. 1987, \araa, 25, 303

\bibitem[{{Delbouille} {et~al.}(1973){Delbouille}, {Roland}, \&
  {Neven}}]{Delbouille+73}
{Delbouille}, L., {Roland}, G., \& {Neven}, L. 1973, {Atlas photometrique DU
  spectre solaire de $\lambda$3000 a $\lambda$10000} (Liege: Universite de
  Liege, Institut d'Astrophysique)

\bibitem[{{Dennison} {et~al.}(1998){Dennison}, {Simonetti}, \&
  {Topasna}}]{VTSLS}
{Dennison}, B., {Simonetti}, J.~H., \& {Topasna}, G.~A. 1998, Publ. Astron.
  Soc. Australia, 15, 147

\bibitem[{{Dettmar}(1998)}]{Dettmar98}
{Dettmar}, R.-J. 1998, Lecture Notes Phys., 506, 527

\bibitem[{{Dickey} \& {Lockman}(1990)}]{DL90}
{Dickey}, J.~M. \& {Lockman}, F.~J. 1990, \araa, 28, 215

\bibitem[{{Domg\"{o}rgen} \& {Dettmar}(1997)}]{DD97}
{Domg\"{o}rgen}, H. \& {Dettmar}, R.-J. 1997, \aap, 322, 391

\bibitem[{{Dove} \& {Shull}(1994)}]{DS94}
{Dove}, J.~B. \& {Shull}, J.~M. 1994, \apj, 430, 222

\bibitem[{{Dove} {et~al.}(2000){Dove}, {Shull}, \& {Ferrara}}]{DSF00}
{Dove}, J.~B., {Shull}, J.~M., \& {Ferrara}, A. 2000, \apj, 531, 846

\bibitem[{{East} \& {Reay}(1984)}]{ER84}
{East}, I.~R. \& {Reay}, N.~K. 1984, \aap, 139, 512

\bibitem[{{Ferri{\` e}re}(2001)}]{Ferriere01}
{Ferri{\` e}re}, K.~M. 2001, Rev. Mod. Phys., 73, 1031

\bibitem[{{Fried}(1978)}]{Fried78}
{Fried}, J.~W. 1978, \aap, 68, 259

\bibitem[{{Gaustad} {et~al.}(2001){Gaustad}, {McCullough}, {Rosing}, \& {Van
  Buren}}]{SHASSA}
{Gaustad}, J.~E., {McCullough}, P.~R., {Rosing}, W., \& {Van Buren}, D. 2001,
  \pasp, 113, 1326

\bibitem[{{Golla} {et~al.}(1996){Golla}, {Dettmar}, \& {Domgoergen}}]{GDD96}
{Golla}, G., {Dettmar}, R.-J., \& {Domgoergen}, H. 1996, \aap, 313, 439

\bibitem[{{Gu{\' e}lin}(1974)}]{Guelin74}
{Gu{\' e}lin}, M. 1974, in IAU Symp. 60, Galactic Radio Astronomy, ed. F.~J.
  Kerr \& S.~C. Simonson (Boston: D. Reidel), 51

\bibitem[{{Haffner}(2001)}]{Haffner01}
{Haffner}, L.~M. 2001, in ASP Conf. Ser. 231, Tetons 4: Galactic Structure,
  Stars and the Interstellar Medium, ed. C.~E. Woodward, M.~D. Bicay, \& J.~M.
  Shull (San Francisco: ASP), 345

\bibitem[{{Haffner} {et~al.}(1998){Haffner}, {Reynolds}, \& {Tufte}}]{HRT98}
{Haffner}, L.~M., {Reynolds}, R.~J., \& {Tufte}, S.~L. 1998, \apjl, 501, L83

\bibitem[{{Haffner} {et~al.}(1999){Haffner}, {Reynolds}, \& {Tufte}}]{HRT99}
---. 1999, \apj, 523, 223

\bibitem[{{Haffner} {et~al.}(2001){Haffner}, {Reynolds}, \& {Tufte}}]{HRT01}
---. 2001, \apjl, 556, L33

\bibitem[{{Hausen} {et~al.}(2002{\natexlab{a}}){Hausen}, {Reynolds}, \&
  {Haffner}}]{Hausen+02-aj}
{Hausen}, N.~R., {Reynolds}, R.~J., \& {Haffner}, L.~M. 2002{\natexlab{a}},
  \aj, 124, 3336

\bibitem[{{Hausen} {et~al.}(2002{\natexlab{b}}){Hausen}, {Reynolds}, {Haffner},
  \& {Tufte}}]{Hausen+02-apj}
{Hausen}, N.~R., {Reynolds}, R.~J., {Haffner}, L.~M., \& {Tufte}, S.~L.
  2002{\natexlab{b}}, \apj, 565, 1060

\bibitem[{{Heiles} {et~al.}(1999){Heiles}, {Haffner}, \& {Reynolds}}]{HHR99}
{Heiles}, C., {Haffner}, L.~M., \& {Reynolds}, R.~J. 1999, in ASP Conf. Ser.
  168, New Perspectives on the Interstellar Medium, ed. A.~R. Taylor, T.~L.
  Landecker, \& G.~Joncas (San Francisco: ASP), 211

\bibitem[{{Heiles} {et~al.}(2000){Heiles}, {Haffner}, {Reynolds}, \&
  {Tufte}}]{Heiles+00}
{Heiles}, C., {Haffner}, L.~M., {Reynolds}, R.~J., \& {Tufte}, S.~L. 2000,
  \apj, 536, 335

\bibitem[{{Hoopes} \& {Walterbos}(2003)}]{HW03}
{Hoopes}, C.~G. \& {Walterbos}, R.~A.~M. 2003, \apj, 586, 902

\bibitem[{{Hoopes} {et~al.}(1999){Hoopes}, {Walterbos}, \& {Rand}}]{HWR99}
{Hoopes}, C.~G., {Walterbos}, R.~A.~M., \& {Rand}, R.~J. 1999, \apj, 522, 669

\bibitem[{{Hoyle} \& {Ellis}(1963)}]{HE63}
{Hoyle}, F. \& {Ellis}, G. R.~A. 1963, Australian J.\ Phys., 16, 1

\bibitem[{{Jahoda} {et~al.}(1990){Jahoda}, {Lockman}, \& {McCammon}}]{JLM90}
{Jahoda}, K., {Lockman}, F.~J., \& {McCammon}, D. 1990, \apj, 354, 184

\bibitem[{{Koo} {et~al.}(1992){Koo}, {Heiles}, \& {Reach}}]{KHR92}
{Koo}, B., {Heiles}, C., \& {Reach}, W.~T. 1992, \apj, 390, 108

\bibitem[{{Lagache} {et~al.}(2000){Lagache}, {Haffner}, {Reynolds}, \&
  {Tufte}}]{Lagache+00}
{Lagache}, G., {Haffner}, L.~M., {Reynolds}, R.~J., \& {Tufte}, S.~L. 2000,
  \aap, 354, 247

\bibitem[{{Levasseur-Regourd} {et~al.}(2001){Levasseur-Regourd}, {Mann},
  {Dumont}, \& {Hanner}}]{LR+01}
{Levasseur-Regourd}, A.~C., {Mann}, I., {Dumont}, R., \& {Hanner}, M.~S. 2001,
  in {Interplanetary Dust}, ed. E.~{Gr\"{u}n}, B.~A.~S. {Gustafson},
  S.~{Dermott}, \& H.~{Fechtig} (New York: Springer), 57

\bibitem[{{Lockman} {et~al.}(1986){Lockman}, {Jahoda}, \& {McCammon}}]{LJM86}
{Lockman}, F.~J., {Jahoda}, K., \& {McCammon}, D. 1986, \apj, 302, 432

\bibitem[{{Madsen} {et~al.}(2002){Madsen}, {Haffner}, \& {Reynolds}}]{MHR02}
{Madsen}, G.~J., {Haffner}, L.~M., \& {Reynolds}, R.~J. 2002, in ASP Conf. Ser.
  276, Seeing Through the Dust: The Detection of HI and the Exploration of the
  ISM in Galaxies, ed. A.~R. Taylor, T.~L. Landecker, \& A.~G. Willis (San
  Francisco: ASP), 96

\bibitem[{{Ma{\'{\i}}z-Apell{\' a}niz}(2001)}]{Maiz-Apellaniz01}
{Ma{\'{\i}}z-Apell{\' a}niz}, J. 2001, \apjl, 560, L83

\bibitem[{{McKee} \& {Ostriker}(1977)}]{MO77}
{McKee}, C.~F. \& {Ostriker}, J.~P. 1977, \apj, 218, 148

\bibitem[{{Miller} \& {Cox}(1993)}]{MC93}
{Miller}, W.~W., I. \& {Cox}, D.~P. 1993, \apj, 417, 579

\bibitem[{{Norman}(1991)}]{Norman91}
{Norman}, C.~A. 1991, in IAU Symp. 144, The Interstellar Disk-Halo Connection
  in Galaxies, ed. H.~Bloemen (Dordrecht: Kluwer), 337--344

\bibitem[{{Nossal} {et~al.}(2001){Nossal}, {Roesler}, {Bishop}, {Reynolds},
  {Haffner}, {Tufte}, {Percival}, \& {Mierkiewicz}}]{Nossal+01}
{Nossal}, S., {Roesler}, F.~L., {Bishop}, J., {Reynolds}, R.~J., {Haffner}, M.,
  {Tufte}, S., {Percival}, J., \& {Mierkiewicz}, E.~J. 2001, \jgr, 106, 5605

\bibitem[{{Nossal} {et~al.}(1998){Nossal}, {Roesler}, \& {Coakley}}]{NRC98}
{Nossal}, S., {Roesler}, F.~L., \& {Coakley}, M.~M. 1998, \jgr, 103, 381

\bibitem[{{Nossal} {et~al.}(1997){Nossal}, {Roesler}, {Coakley}, \&
  {Reynolds}}]{Nossal+97}
{Nossal}, S., {Roesler}, F.~L., {Coakley}, M.~M., \& {Reynolds}, R.~J. 1997,
  \jgr, 102, 14541

\bibitem[{{Otte} {et~al.}(2002){Otte}, {Gallagher}, \& {Reynolds}}]{OGR02}
{Otte}, B., {Gallagher}, J.~S., \& {Reynolds}, R.~J. 2002, \apj, 572, 823

\bibitem[{{Otte} {et~al.}(2001){Otte}, {Reynolds}, {Gallagher}, \&
  {Ferguson}}]{Otte+01}
{Otte}, B., {Reynolds}, R.~J., {Gallagher}, J.~S., \& {Ferguson}, A.~M.~N.
  2001, \apj, 560, 207

\bibitem[{{Parker} \& {Phillipps}(1998)}]{UKST}
{Parker}, Q.~A. \& {Phillipps}, S. 1998, Publ. Astron. Soc. Australia, 15, 28

\bibitem[{{Percival}(1995)}]{Percival95}
{Percival}, J.~W. 1995, in Proc. SPIE Vol. 2479, Telescope Control Systems, ed.
  P.~T. Wallace, 33

\bibitem[{{Rand}(1996)}]{Rand96}
{Rand}, R.~J. 1996, \apj, 462, 712

\bibitem[{{Rand}(1997)}]{Rand97}
---. 1997, \apj, 474, 129

\bibitem[{{Rand}(1998)}]{Rand98}
---. 1998, \apj, 501, 137

\bibitem[{{Rand}(2000)}]{Rand00}
---. 2000, \apjl, 537, L13

\bibitem[{{Reynolds}(1990)}]{Reynolds90}
{Reynolds}, R.~J. 1990, in IAU Symp. 139, The Galactic and Extragalactic
  Background Radiation, ed. S.~Bowyer \& C.~Leinert (Dordrecht: Kluwer),
  157--169

\bibitem[{{Reynolds}(1993)}]{Reynolds93}
{Reynolds}, R.~J. 1993, in AIP Conf. Proc. 278, Back to the Galaxy, ed. S.~S.
  Hold \& F.~Verter (New York: AIP), 156

\bibitem[{{Reynolds}(1997)}]{Reynolds97}
{Reynolds}, R.~J. 1997, in The Physics of Galactic Halos, ed. H.~Lesch, R.-J.
  Dettmar, U.~Mebold, \& R.~Schlickeiser (Berlin: Akademie), 57

\bibitem[{{Reynolds} {et~al.}(1999){Reynolds}, {Haffner}, \& {Tufte}}]{RHT99}
{Reynolds}, R.~J., {Haffner}, L.~M., \& {Tufte}, S.~L. 1999, \apjl, 525, L21

\bibitem[{{Reynolds} {et~al.}(1998){Reynolds}, {Hausen}, {Tufte}, \&
  {Haffner}}]{Reynolds+98}
{Reynolds}, R.~J., {Hausen}, N.~R., {Tufte}, S.~L., \& {Haffner}, L.~M. 1998,
  \apjl, 494, L99

\bibitem[{{Reynolds} {et~al.}(1990){Reynolds}, {Roesler}, {Scherb}, \&
  {Harlander}}]{RRSH90}
{Reynolds}, R.~J., {Roesler}, F.~L., {Scherb}, F., \& {Harlander}, J. 1990, in
  Instrumentation in astronomy VII, ed. D.~Crawford (Bellingham: SPIE),
  610--621

\bibitem[{{Reynolds} {et~al.}(2001{\natexlab{a}}){Reynolds}, {Sterling}, \&
  {Haffner}}]{RSH01}
{Reynolds}, R.~J., {Sterling}, N.~C., \& {Haffner}, L.~M. 2001{\natexlab{a}},
  \apjl, 558, L101

\bibitem[{{Reynolds} {et~al.}(2001{\natexlab{b}}){Reynolds}, {Sterling},
  {Haffner}, \& {Tufte}}]{Reynolds+01}
{Reynolds}, R.~J., {Sterling}, N.~C., {Haffner}, L.~M., \& {Tufte}, S.~L.
  2001{\natexlab{b}}, \apjl, 548, L221

\bibitem[{{Robley} {et~al.}(1985){Robley}, {Buecher}, {Koutchmy}, \&
  {Lamy}}]{Robley+85}
{Robley}, R., {Buecher}, A., {Koutchmy}, S., \& {Lamy}, P. 1985, in ASSL Vol.
  119: IAU Colloq. 85: Properties and Interactions of Interplanetary Dust, ed.
  R.~Giese \& P.~Lamy (Dordrecht: D. Reidel), 85--88

\bibitem[{{Rossa}(2001)}]{RossaPhD}
{Rossa}, J. 2001, Ph.D.~Thesis

\bibitem[{{Rossa} \& {Dettmar}(2000)}]{RD00}
{Rossa}, J. \& {Dettmar}, R.-J. 2000, \aap, 359, 433

\bibitem[{{Scherb}(1981)}]{Scherb81}
{Scherb}, F. 1981, \apj, 243, 644

\bibitem[{{Sciama}(1990)}]{Sciama90}
{Sciama}, D.~W. 1990, \apj, 364, 549

\bibitem[{{Slavin} {et~al.}(2000){Slavin}, {McKee}, \& {Hollenbach}}]{SM00}
{Slavin}, J.~D., {McKee}, C.~F., \& {Hollenbach}, D.~J. 2000, \apj, 541, 218

\bibitem[{{Smith} \& {Cox}(2001)}]{SC01}
{Smith}, R.~K. \& {Cox}, D.~P. 2001, \apjs, 134, 283

\bibitem[{{Spitzer} \& {Fitzpatrick}(1993)}]{SF93}
{Spitzer}, L.~J. \& {Fitzpatrick}, E.~L. 1993, \apj, 409, 299

\bibitem[{{T{\" u}llmann} \& {Dettmar}(2000)}]{TD00}
{T{\" u}llmann}, R. \& {Dettmar}, R.-J. 2000, \aap, 362, 119

\bibitem[{{Thilker} {et~al.}(2002){Thilker}, {Walterbos}, {Braun}, \&
  {Hoopes}}]{Thilker+02}
{Thilker}, D.~A., {Walterbos}, R.~A.~M., {Braun}, R., \& {Hoopes}, C.~G. 2002,
  \aj, 124, 3118

\bibitem[{Tufte(1997)}]{TuftePhD}
Tufte, S.~L. 1997, PhD thesis, University of Wisconsin--Madison

\bibitem[{{Tufte} {et~al.}(1998){Tufte}, {Reynolds}, \& {Haffner}}]{TRH98}
{Tufte}, S.~L., {Reynolds}, R.~J., \& {Haffner}, L.~M. 1998, \apj, 504, 773

\bibitem[{{Tufte} {et~al.}(2002){Tufte}, {Wilson}, {Madsen}, {Haffner}, \&
  {Reynolds}}]{Tufte+02}
{Tufte}, S.~L., {Wilson}, J.~D., {Madsen}, G.~J., {Haffner}, L.~M., \&
  {Reynolds}, R.~J. 2002, \apjl, 572, L153

\bibitem[{{Wakker}(2001)}]{Wakker01}
{Wakker}, B.~P. 2001, \apjs, 136, 463

\bibitem[{{Wakker} {et~al.}(1999){Wakker}, {Howk}, {Savage}, {van Woerden},
  {Tufte}, {Schwarz}, {Benjamin}, {Reynolds}, {Peletier}, \&
  {Kalberla}}]{Wakker+99}
{Wakker}, B.~P., {Howk}, J.~C., {Savage}, B.~D., {van Woerden}, H., {Tufte},
  S.~L., {Schwarz}, U.~J., {Benjamin}, R., {Reynolds}, R.~J., {Peletier},
  R.~F., \& {Kalberla}, P. M.~W. 1999, \nat, 402, 388

\bibitem[{{Wood} \& {Reynolds}(1999)}]{WR99}
{Wood}, K. \& {Reynolds}, R.~J. 1999, \apj, 525, 799

\bibitem[{{Zurita} {et~al.}(2002){Zurita}, {Beckman}, {Rozas}, \&
  {Ryder}}]{Zurita+02}
{Zurita}, A., {Beckman}, J.~E., {Rozas}, M., \& {Ryder}, S. 2002, \aap, 386,
  801

\bibitem[{{Zurita} {et~al.}(2000){Zurita}, {Rozas}, \& {Beckman}}]{ZRB00}
{Zurita}, A., {Rozas}, M., \& {Beckman}, J.~E. 2000, \aap, 363, 9

\end{thebibliography}



\clearpage

\begin{deluxetable}{cccc}
\tablewidth{0pt}
\tablecaption{Faint Atmospheric Lines near \ha\label{tab:atmlines}}
\tablehead{
 & \colhead{$v_\mathrm{geo}$} & \colhead{FWHM} & \colhead{Relative} \\
\colhead{Line} & \colhead{(\kms)} & \colhead{(\kms)} & \colhead{Intensity}
}

\startdata
1 & $-148.5$ & 10 & 0.54  \\
2 & $-131.6$ & 10 & 0.24  \\
3 & $-88.6$ & 10 & 0.20  \\
4 & $-61.7$ & 15 & 0.33  \\
5 & $-42.0$ & 10 & 1.00  \\
6 & $-25.1$ & 10 & 0.30  \\
7 & $+32.4$ & 10 & 0.30  \\
8 & $+40.7$ & 10 & 0.65  \\
9 & $+53.9$ & 10 & 0.39  \\
10 & $+72.6$ & 15 & 1.60  \\
11 & $+95.9$ & 10 & 0.36  \\
12 & $+120.6$ & 10 & 0.29  \\
\enddata

\end{deluxetable}

\clearpage

\begin{deluxetable}{cccccc}
\tablewidth{0pt}
\tablecaption{Integrated \ha\ Emission between \vlsr $= -80$ and $+80$ \kms
\label{table:total}}
\tablehead{ 
  \multicolumn{2}{c}{Position} & \colhead{\iha} &
    \colhead{$\sigma$\tablenotemark{b}} &
    \colhead{Block\tablenotemark{c}} & \colhead{Original\tablenotemark{d}} \\
  \cline{1-2}
  \colhead{$\ell$} & \colhead{$b$} & \colhead{(R\tablenotemark{a})} & 
    \colhead{(R)} & \colhead {} & \colhead{\iha (R)}
  }

\startdata
    $76\fdg189$  & $-46\fdg682$  &   1.204  &   0.000  &    1000  &   1.204 \\
    $77\fdg619$  & $-46\fdg682$  &   1.285  &   0.034  &    1000  &   0.000 \\
    $79\fdg049$  & $-46\fdg682$  &   1.047  &   0.035  &    1000  &   0.000 \\
    $80\fdg479$  & $-46\fdg682$  &   0.700  &   0.034  &    1000  &   0.000 \\
    $81\fdg569$  & $-45\fdg832$  &   0.668  &   0.034  &    1000  &   0.000 \\
    $80\fdg169$  & $-45\fdg832$  &   0.634  &   0.037  &    1000  &   0.000 \\
    $78\fdg759$  & $-45\fdg832$  &   0.867  &   0.035  &    1000  &   0.000 \\
    $77\fdg349$  & $-45\fdg832$  &   1.205  &   0.035  &    1000  &   0.000 \\
    $75\fdg949$  & $-45\fdg832$  &   0.993  &   0.035  &    1000  &   0.000 \\
    $75\fdg309$  & $-44\fdg981$  &   1.108  &   0.034  &    1000  &   0.000 \\
    \multicolumn{6}{c}{$\vdots$} \\
\enddata

\tablenotetext{a}{1 R $= 10^{6}/4\pi$ photons cm$^{-2}$ s$^{-1}$ sr$^{-1} = 2.4 \times 10^{-7}$ ergs cm$^{-2}$ s$^{-1}$ sr$^{-1}$ at \ha.}
\tablenotetext{b}{Formal propagation only. This error does not reflect systematic issues discussed in the text. If listed as zero, the original observation is likely contaminated by a bright star (see text). The intensity listed in the third column (\iha) is then an average of the nearest neighbors within one degree of the original.}
\tablenotetext{c}{The numeric block designation for this observation. Blocks of data are taken sequentially in time and are physically close on the sky (see text). Systematics remaining in the data are thus often linked to this parameter.}
\tablenotetext{d}{If the third column represents the average of nearest neighbors (i.e. the fourth column is zero; see note \emph{b} above), this column contains the original (unmodified) intensity for reference.}

\tablecomments{The complete version of this table is in the electronic edition of
the Journal. The printed edition contains only a sample.}

\end{deluxetable}

\clearpage

\end{document}